\documentclass{sig-alternate}
\usepackage{times}
\usepackage{subfigure}
\usepackage{graphicx}
\usepackage{amsmath}
\usepackage{bm}
\usepackage{url}
\usepackage{stmaryrd}
\usepackage{balance}
\usepackage{amssymb}
\usepackage{pgfplots}
\usepackage{pifont}
\usepackage[usenames,dvipsnames]{pstricks}
\usepackage{epsfig}
\usepackage{booktabs}
\usepackage{pgffor}
\usepackage{tikz}
\usepackage{multirow}
\usepackage{algorithm}
\usepackage{algorithmic}
\definecolor{purple}{rgb}{0.6,0.2,0.8}

\newdef{definition}{Definition}

\newcommand{\mb}{\mathbf}

\usepackage{algorithm}
\usepackage{algorithmic}

\begin{document}
\title{Badge System Analysis and Design}
\numberofauthors{3}
\author{
\alignauthor
Jiawei~Zhang\\
      \affaddr{University of Illinois at Chicago}\\
      \affaddr{Chicago, IL, USA}\\
       \email{jzhan9@uic.edu}
\alignauthor
Xiangnan~Kong\\
       \affaddr{Worcester Polytechnic Institute}\\
       \affaddr{Worcester, MA, USA}\\
       \email{xkong@wpi.edu}
\alignauthor
Philip~S.~Yu\\
       \affaddr{University of Illinois at Chicago}\\
       \affaddr{Chicago, IL, USA}\\
       \email{psyu@cs.uic.edu}
}

\maketitle

\begin{abstract}

To incentivize users' participations and steer their online activities, online social networks start to provide users with various kinds of rewards for their contributions to the sites. The most frequently distributed rewards include account levels, reputation scores, different kinds of badges, and even material awards like small gifts and cash back, etc. Attracted by these rewards, users will spend more time using the network services. In this paper, we will mainly focus on ``badges reward systems'' but the proposed models can be applied to other reward systems as well.

Badges are small icons attached to users' homepages and profiles denoting their achievements. People like to accumulate badge for various reasons and different badges can have specific \textit{values} for them. Meanwhile, to get badges, they also need to exert efforts to finish the required tasks, which can lead to certain \textit{costs}. To understand and model users' motivations in badge achievement activities, we will study an existing badge system launched inside a real-world online social network, Foursquare, in this paper. 

At the same time, to maximize users' contributions to online social networks, social network \textit{system designers} need to determine the optimal badge system mechanism carefully. Badge system mechanism describes various detailed aspects of the system and can involve many parameters, e.g., categories of existing badges, number of badges available as well as the minimum contributions required to obtain the badges, which all need to be designed with meticulous investigations. Based on the model of users' badges accumulating activities, in this paper, we will also study how to design the badge system that can incentivize the maximum users' contributions to the social networks.

\end{abstract}

\begin{figure}[t]
\centering
\subfigure[shared badge distribution]{ \label{eg_fig_1_2}
    \begin{minipage}[l]{0.47\columnwidth}
      \centering
      \includegraphics[width=1.0\textwidth]{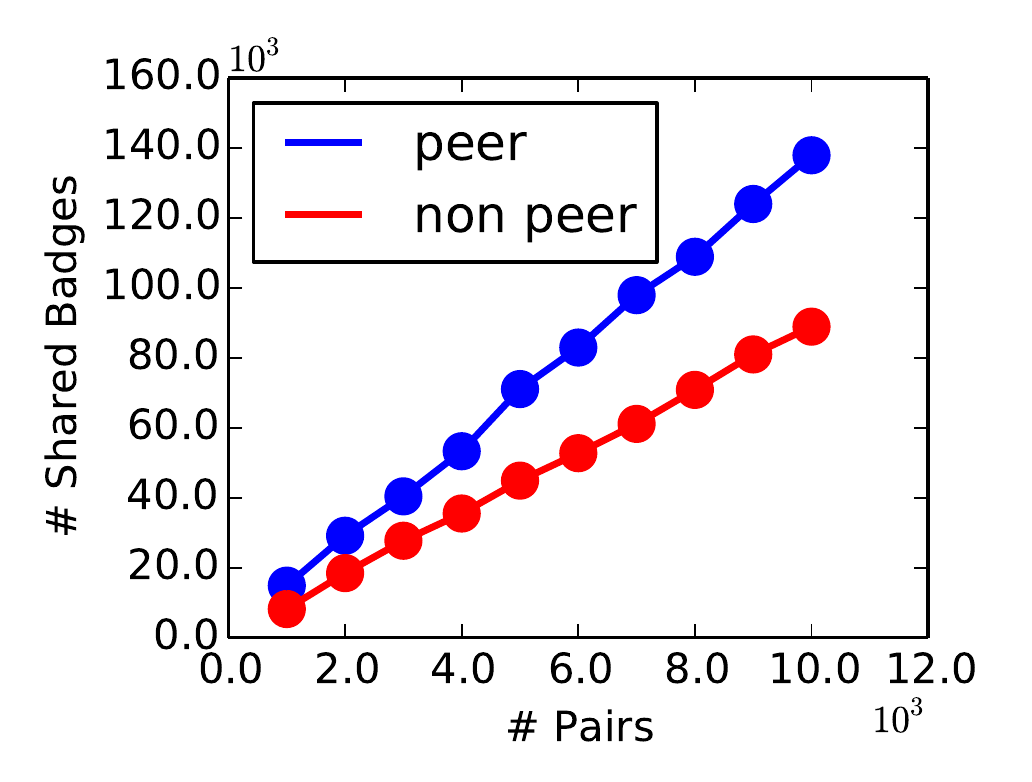}
    \end{minipage}
}
\subfigure[peer pressure distribution]{\label{eg_fig_1_3}
    \begin{minipage}[l]{0.47\columnwidth}
      \centering
      \includegraphics[width=1.0\textwidth]{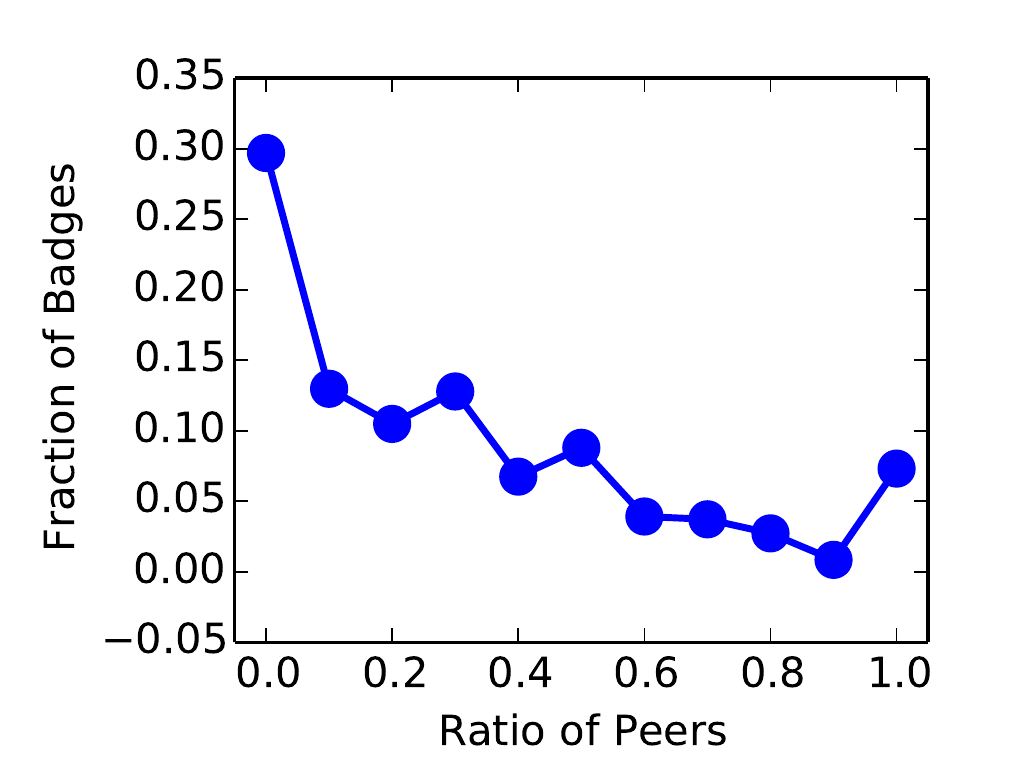}
    \end{minipage}
}
\vspace{-8pt}
\caption{Statistical information: (a) \#shared badges between pairs in a randomly sampled pair set; (b) fraction of badges obtained at a certain friend ratio.}\label{eg_fig_1}\vspace{-8pt}
\end{figure}

\begin{figure}[t]
\centering
    \begin{minipage}[l]{0.9\columnwidth}
      \centering
      \includegraphics[width=\textwidth]{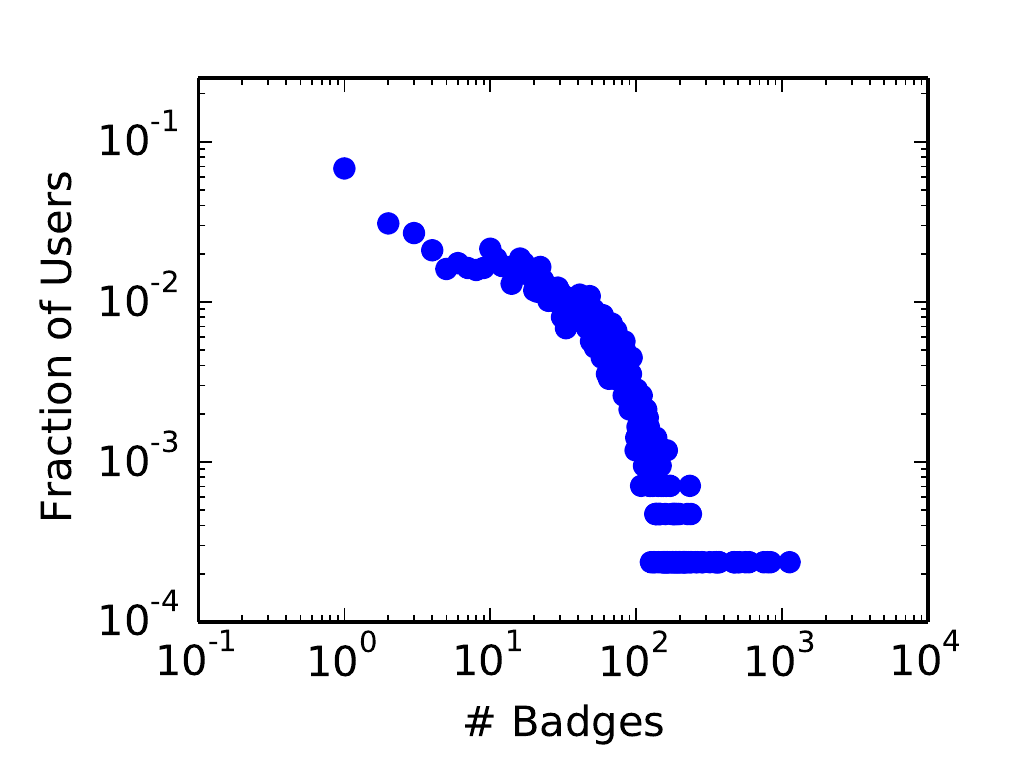}
    \end{minipage}
\vspace{-8pt}
  \caption{Power law distribution of \textit{user fraction} and \textit{number of achieved badges}.}\label{fig:badge_power}\vspace{-8pt}
\end{figure}


\begin{table}[t]
\caption{Properties of the Badge System Dataset}
\label{tab:datastat}
\centering
\begin{tabular}{clr}
\toprule

&property &number \\
\midrule 
\multirow{2}{*}{nodes}
&user		& 4,240 \\
&badge	& 1,431 \\
\midrule 
\multirow{3}{*}{links}
&follow		&81,291 \\
&achieved		& 176,301\\
&level		&47,342\\
\bottomrule
\end{tabular}\vspace{-10pt}
\end{table}


\section{Introduction to Badge System}\label{sec:intro}


Social networks, e.g., Facebook, Twitter and Foursquare, have achieved remarkable success in recent years. These social networks are mostly driven by user-generated content, e.g., posts, photos and location checkins. To incentivize users' participations and steer their online activities, many social networks start to offer users various rewards for their contributions to the networks. In this paper, we will mainly focus on ``badge reward systems'' but the proposed models can be applied to other reward systems as well. Badge systems have been adopted by a wide range of social networks: (1) Foursquare\footnote{https://foursquare.com}, a famous location-based social network (LBSN), is distributing different badges to users for their geo-location checkins; (2) Weibo\footnote{http://www.weibo.com}, a social media in China, launches a badge system to give users badges for writing posts and replies; (3) Stack Overflow\footnote{http://stackoverflow.com}, a popular question and answer (Q\&A) site for computer programming, adopts a system where users can get badges by answering questions in the site; and (4) In Khan Academy\footnote{https://www.khanacademy.org}, a popular massive open online course (MOOC) site, users are awarded badges for watching course videos and answering questions.

\subsection{User Activity Observations}

People like to accumulate badges for various reasons and different badges can have specific \textit{values} for them. Extensive analyses have been done on a real-world badge system launched in Foursquare. The badge system dataset was crawled from Foursquare during the April of 2014. We collected $4,240$ users together with all the $1,430$ categories of badges achieved by them. These users are crawled with BFS search from several random seed users via the social connections in Foursquare, who are connected by \textit{follow} links of number $81,291$. To denote that a user has achieved certain badges, we add \textit{achieve} links between users and badges, whose total number is $176,301$ in the crawled dataset. On average, each user has achieved $42$ badges in Foursquare. In addition, each category of badges can involve different badge levels, where badges of consequential levels can be connected by the \textit{level} links. For example, badges of higher level, e.g., $l$ ($l > 1$), can have \textit{level} links pointing to badge of level $l - 1$ and the number of \textit{level} links among badges is $47,342$. A more detailed information about the dataset is available in Table~\ref{tab:datastat}.

The statistical analyses results about the badge system dataset are available in Figure~\ref{eg_fig_1} and Table~\ref{tab:top_badge}, from which many interesting phenomena can be observed:


\begin{enumerate}

\item \textit{Generally, users who are friends are more likely to share common badges}. We randomly sample a certain number of user pairs who are (1) friends (i.e., connected by social links) and (2) not friends from Foursquare, and count the number of common badges shared by these user pairs respectively. The results are given in Figure~\ref{eg_fig_1_2}, where the x axis is the number of randomly sampled user pairs and the y axis denotes the number of shared badges between these sampled pairs. From Figure~\ref{eg_fig_1_2}, we can observe that online badge achievement in social networks is correlated with social connections among users and friends are more likely to have common badges.

\item \textit{In many cases, users are likely to obtain badges which have never been achieved by his friends}. As shown in Figure~\ref{eg_fig_1_3}, for each badge $b_j$, obtained by user $u_i$, we get the timestamp when $u_i$ get $b_j$ and the ratio of $u_i$'s friends who obtain $b_j$ before $u_i$. The distribution of the percentage of badges obtained at different ratios is given in Figure~\ref{eg_fig_1_3}, from which we can observe that a large proportion of badges are obtained at small ratios, i.e., few of $u_i$'s friends have achieved the badge before $u_i$.

\begin{figure}[t]
\centering
    \begin{minipage}[l]{1.0\columnwidth}
      \centering
      \includegraphics[width=\textwidth]{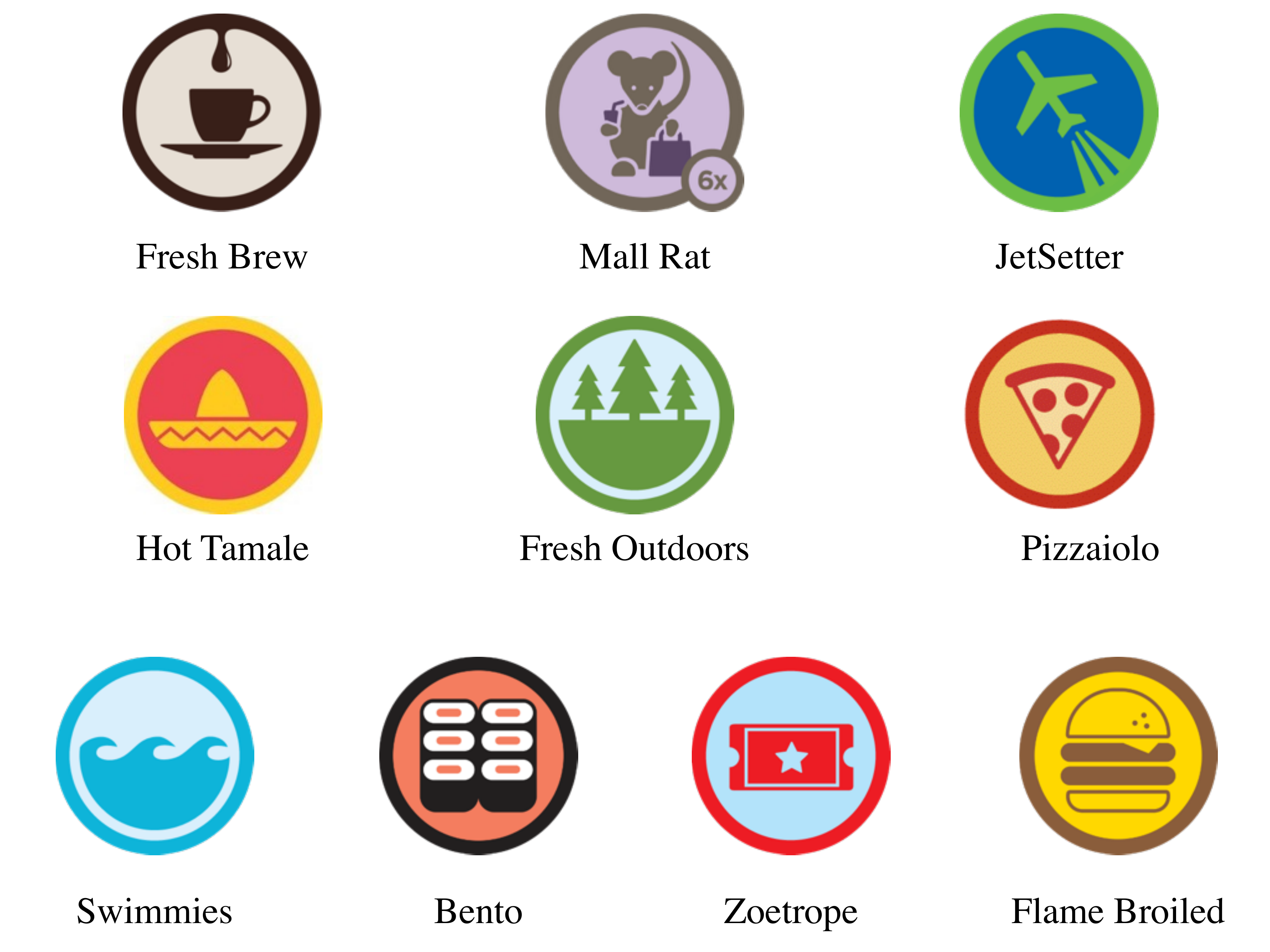}
    \end{minipage}
\vspace{-8pt}
  \caption{Top 10 badges achieved by most users}\label{fig:badge_list}\vspace{-8pt}
\end{figure}

\begin{figure}[t]
\centering
    \begin{minipage}[l]{1.0\columnwidth}
      \centering
      \includegraphics[width=\textwidth]{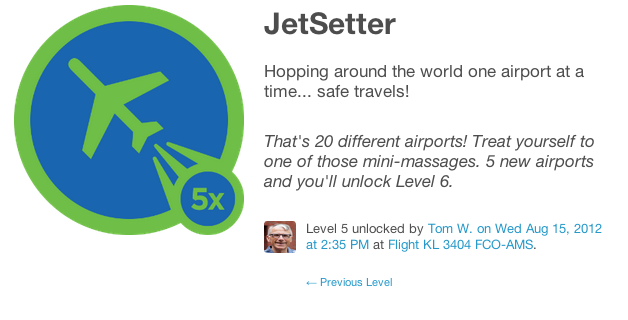}
    \end{minipage}
\vspace{-8pt}
  \caption{Tasks needed to unlock ``JetSetter'' badges.}\label{fig:jetsetter_unlock}\vspace{-8pt}
\end{figure}


\begin{table*}[t]
\caption{Number of users achieving top 10 badges}
\label{tab:top_badge}
\centering
\begin{tabular}{ccccccccccccc}
\toprule

badge & obtain it by &\multicolumn{10}{c}{\# users achieving badges of different levels}  &total  \\
\cmidrule{3-12}
name &checking-in at &1 &2 &3 &4 &5 &6 &7 &8 &9 &10 &number\\
\midrule
Fresh Brew &Coffee Shops     &2468     &1914     &1235     &817     &555     &374     &255     &144     &78     &38     &7878\\
Mall Rat &Shopping Malls    &2545     &1907     &1076     &624     &366     &224     &130     &81     &46     &29     &7028\\
JetSetter &Airport Terminals    &2357     &1703     &972     &564     &339     &210     &147     &102     &63     &11     &6468\\
Hot Tamale  &Mexican Restaurants   &2305     &1733     &989     &583     &336     &191     &105     &58     &37     &18     &6355\\
Great Outdoors  &Parks and Outdoors   &2119     &1535     &801     &468     &295     &200     &132     &95     &53     &30     &5728\\
Pizzaiolo   &Pizza Restaurants  &2192     &1450     &605     &267     &116     &62     &26     &16     &8     &4     &4746\\
Swimmies  &Lake/Pond/Beach   &1888     &1214     &538     &281     &159     &107     &74     &47     &36     &17     &4361\\
Bento   &Sushi Restaurants  &1741     &1121     &459     &209     &104     &63     &34     &21     &14     &8     &3774\\
Zoetrope  &Movie Theaters   &1985     &1106     &309     &103     &34     &16     &12     &6     &5     &4     &3580\\
Flame Broiled   &Burger Restaurants   &1944     &1044     &337     &105     &40     &13     &6     &3     &1     &1     &3494\\

\bottomrule
\end{tabular}\vspace{-10pt}
\end{table*}


\item \textit{In some cases, users will follow their friends when most of their peers have obtained a certain badge}. Still in Figure~\ref{eg_fig_1_3}, when the ratio is close to $1.0$, i.e., almost all the peers have obtained a certain badge, the fraction of badge obtained will increase to $0.1$, representing that about $10\%$ of the badges are obtained by users when all his friends have achieved the badge. As a result, users' badge achievement activities in online social networks are highly related to those of their peers. These three observed effects of peers' badges on users' activities are called the \textit{peer leadership value} of badges in Section~\ref{subsec:peer}.

\item \textit{Users are keen on getting badges that are to their interests}. In Table~\ref{tab:top_badge}, we extract top $10$ popular badges achieved by the most users in Foursquare and each of these badges has $10$ different levels. Numbers of users who have obtained certain levels of each kind of badge as well as the total number of badges achieved (i.e., summation of each row of the table) are provided. The corresponding icons of these badges are shown in Figure~\ref{fig:badge_list}. Generally, higher-level badges require more efforts from the users, but from Table~\ref{tab:top_badge}, we observe there are still a large number of users are willing to devote such high efforts to get the badges due to their personal interests. For example, among the $2,468$ users who achieved the ``Fresh Brew'' badge of level 1, $22.5\%$ of them will continue to get badge of level 5, which may denote these users like drinking coffee a lot. Similarly, for users who get different levels of ``Mall Rat'' badges, they should like shopping a lot; users who get the ``JetSetter'' badges are those who travel frequently. As a result, badges can reveal users personal interests, especially higher-level badges. Such a kind of effects of badges for users is defined as the \textit{personal interest value} of badges in Section~\ref{subsec:interest}.


\item \textit{Users in online social networks are enthusiastic in earning badges}. As shown in Figure~\ref{fig:badge_power}, the distribution of user fractions obtaining a certain number badges follows the \textit{power law} \cite{CSN09} and the majority of the users in Foursquare have obtained more than $50$ badges in Foursquare. In Table~\ref{tab:top_badge}, we observe that over $2,000$ have ever obtained the ``Fresh Brew'', ``Mall Rat'', ``JetSetter'', ``Hot Tamale'', ``Great Outdoors'' and ``Pizzaiolo'' badges. Considering that there are only $4,240$ users in the dataset, we can observe that most users in Foursquare have achieved these badges. One potential explanation of the observation can be that: initially, people use social network services independently of the badges; meanwhile, after using the network for a while, users start to lose interesting, and the badges start to play their roles in keeping users within the network. Such a kind of effects of badges on users is defined as the \textit{network trend value} of badges in Section~\ref{subsec:network}.


\end{enumerate}

\subsection{Badge System Analysis}

To get badges, users in online social networks are required to finish certain tasks, which can be (1) finishing a certain number of checkins at required locations in Foursquare, (2) answering a number of questions proposed by other users in Stack Overflow, and (3) publishing required numbers of posts in Weibo. These tasks of higher-level badges are usually more challenging. For instance, as shown in Figure~\ref{fig:jetsetter_unlock}, by checking at $20$ different airports, users can get the ``JetSetter'' badges from level 1 to level  5. However, to unlock the ``JetSetter'' badge of level 6, users need to check in at another $5$ new airports. Obviously, these tasks can yield some costs, which can be time, money or knowledge spent on the tasks. The payoff of achieving certain badges is defined as ``(value - cost)'', i.e., the utility of badges for users. Besides the above badge values, users can also get other benefits from the network, e.g., enjoy the social network services. With the same model to be proposed in this paper, such benefits can be handled by either incorporating them into the above corresponding badge value categories or introducing a new value category. Since we mainly focus on the badges system design problem in this paper, we will only consider the above badge values to simplify the problem settings. When the value of badges can exceed the cost, users may try to get the badge; otherwise, they will not devote their efforts as they can get no payoffs from these badges. Costs of obtaining badges are fixed but the value of badges can be influenced by other users' badge achievement activities. Each user in online social network is assumed to be ``selfish'' and wants to maximize his payoff (i.e., the utility) and the badge achieving activities will form a game with other users in the network.



Meanwhile, badges in online social networks are used to incentivize users' contributions to the sites. To achieve such a goal, the system designer needs to determine the optimal settings for a badge system, which is formally defined as the \textit{badge system mechanism} in this paper. Badge system mechanism covers every aspect of the system and contains many parameters, which include:


\begin{itemize}


\item \textit{Categories of Badges}: A badge is said to be \textit{contributing} if it can attract lots of users' contributions to the social network. We will study the contributions attracted by different categories of badges from users and find the most contributing ones in online social networks. This problem is helpful for badge designer to determine which categories of badges should be placed in social networks.

\item \textit{Number of Badges}: The number of badges to be placed in online social networks is very important for an effective badge system. If a large number of badges are placed in social networks, users can always get badges to their interests or have never be obtained by others before very easily, which will make the system fail to work in steering users' online activities. 

\item \textit{Badge Threshold}: Setting the badge contribution threshold (i.e., the minimum required contribution to get the badge) is very tricky. If the threshold is set too low, most users will get the badge easily with little efforts and the total amount of contributions drawn by these badges will be low. On the other hand, if the threshold is set too high, very few people will devote their effort to getting the badge as these badges are difficult to obtain and the total amount of users' contributions will be low as well. 

\end{itemize}

\subsection{Badge System Modeling Challenges}

The \textit{badge system analysis and design} problem studied in this paper is a new research problem and it is very challenging to solve due to the following reasons:

\begin{itemize}

\item \textit{Badge Value Definition}: The value of badges for users in online social networks is unclear. Formal definition, quantification and inference of the badge values for users can be the prerequisite for a comprehensive modeling users' badge achievement activities.

\item \textit{User Utility Function}: Users can get reward from the achieved badges, i.e., the \textit{value} of badges, but also needs devote their efforts, i.e., the cost. Formal definition of the payoff by achieving badges, i.e., the utility, is still an open problem.

\item \textit{Game Among Users}: Each user aims at maximizing his overall utility, which can be influenced by other users' activities at the same time. As a result, users' online badge achieving activities can lead to a game involving all users. Formulation and analysis of such a game among users in online social networks is very difficult.

\item \textit{Optimal Badge Mechanism}: Users' utility in badge achievement is determined by the badge system mechanism set by the designer. Users aim at maximizing their overall utility with as few efforts as possible, while the designer aims at maximizing users' overall contribution to the network. As a result, there also exists a game between users and the designer, which will make the problem more challenging.



\end{itemize}


The following parts of this paper are organized as follows. Section~\ref{sec:concepts} is about the definitions of many important concepts. Various \textit{value functions} of badges for users will be introduced and combined in the \textit{comprehensive value function} in Section~\ref{sec:value}. Users' \textit{utility function} is given in Section~\ref{sec:utility}. In Section~\ref{sec:game1}, we study the game among users in online social network. In Section~\ref{sec:game2}, we solve the \textit{badge system design} problem by formulating it as a game between users and the designer and provide basic simulation analysis about various aspects in the \textit{badge system design} problem. The related works are introduced in Section~\ref{sec:relatedwork}. Finally, we conclude the paper in Section~\ref{sec:conclusion}.

\section{Terminology Definition} \label{sec:concepts}





Users in social networks can be gifted in different areas and they can finish the tasks required to get badges corresponding to their gifts effortlessly. For instance, in Foursquare, sports enthusiasts can get \textit{Gym Rat} badges easily as they do sports in gyms regularly, while travel lovers can obtain \textit{JetSetter} or \textit{Trainspotter} badges by checking in at train stations and airports frequently. However, for users who want to get badges of areas that they are not good at, it would be very difficult to finish the required tasks. For example, a sports enthusiast may need to spend lots of time and money to get the \textit{JetSetter} or \textit{Trainspotter} badges by travelling. Similarly, gourmets who seldom do sports may suffer a lot to get \textit{Gym Rat} badges by visiting gyms. Let $\mathcal{U} = \{u_1, u_2, \cdots, u_n\}$ and $\mathcal{B} = \{b_1, b_2, \cdots, b_m\}$ be the sets containing $n$ users and $m$ badges respectively in the network. To depict such phenomena, we formally define the concepts of \textit{ability}, \textit{effort} and \textit{contribution} of users in $\mathcal{U}$ as well as \textit{contribution threshold} of badges in $\mathcal{B}$ as follows.


\begin{definition}
(Ability): User $u_i$'s talents or advantages in fields corresponding to badges in $\mathcal{B}$ can be represented as the \textit{ability} vector $\mb{a}_i = [a_{i,1}, a_{i,2}, \cdots, a_{i,m}]$, where $a_{i,j} \ge 0$ denotes $u_i$'s \textit{ability} in the field of badge $b_j$ or simply $u_i$'s \textit{ability} to get badge $b_j$.
\end{definition}


All people are assumed to be created equally talented. People can be talented at different aspects but the total intelligence each people have can be quite close. For simplicity, we assume the total abilities of different users are equal, i.e., $\left | \mb{a}_i \right |_1 = \left | \mb{a}_j \right |_1$, for $\forall u_i, u_j \in \mathcal{U}$. Besides talents, to make achievements in certain areas, every people need to devote their efforts and passion, which can be either money, time, energy or knowledge. In this paper, the resources users will devote to the system is time and the formal definition of \textit{unit time effort} is available as follows.

\begin{definition}
(Unit Time Effort): Vector $\mb{e}_i = [e_{i,1}, e_{i,2}, \cdots, e_{i,m}]$ denotes user $u_i$'s efforts devoted to the field corresponding badges in $\mathcal{B}$ in unit time, where $e_{i,j} \ge 0$ represents $u_i$'s \textit{effort} devoted in the area of badge $b_j$, or simply $u_i$'s \textit{effort} in getting badge $b_j$. 
\end{definition}

Users' \textit{unit time effort} can vary with time and can be represented as a function on time, e.g., $e_{i,j}(t)$. The total amount of \textit{unit time effort} in different areas of all users are assumed to be equal, i.e., $\left | \mb{e}_i \right |_1 = \left | \mb{e}_j \right |_1$, for $\forall u_i, u_j \in \mathcal{U}$. Meanwhile, the more time people devote to certain area, the more \textit{cumulative efforts} he will devote to the area.

\begin{definition}
(Cumulative Effort): Term $\hat{e}_{i,j} = \int_{\underline{t}}^{\overline{t}} e_{i,j}(t)\, dt$ is defined as the \textit{cumulative effort} that user $u_i$ devotes to badge $b_j$ during time period $[\underline{t}, \overline{t}]$. For both users and designer, \textit{cumulative effort} is more meaningful as they only care about the total amount of effective efforts devoted to the system. Vector $\mb{\hat{e}}_i = [\hat{e}_{i,1}, \hat{e}_{i,2}, \cdots, \hat{e}_{i,m}]$ is defined as the \textit{cumulative efforts} that user $u_i$ pays to the network. 
\end{definition}

In this paper, active users are assumed to have more \textit{cumulative efforts} as they spend more time using the social network. The achievements people obtain depend on not only their ability in a certain area but also the efforts the devoted to the area, which can be formally defined as their \textit{contributions} to the network.

\begin{definition}
(User Contribution): The effectiveness of users' cumulative efforts devoted to a social network is formally defined as their \textit{contributions}. Vector $\mb{c}_i = [c_{i,1}, c_{i,2}, \cdots, c_{i,m}]$ is defined to be user $u_i$'s contributions to the whole system, where $c_{i,j}$ is the contribution of user $u_i$ devoted to the network in getting badge $b_j$ during $[\underline{t}, \overline{t}]$:
\begin{align*}
c_{i,j} &= \int_{\underline{t}}^{\overline{t}} a_{i,j}e_{i,j}(t)\, dt =a_{i,j} \int_{\underline{t}}^{\overline{t}} e_{i,j}(t)\, dt =a_{i,j} \hat{e}_{i,j}.
\end{align*}
\end{definition}

As a result, the more effort people devote to areas they are gifted at, the more remarkable achievements they can get in the areas. In social networks, whether a user can receive a badge depends on not only the contributions he/she make but also the contribution \textit{threshold} of the badge.

\begin{definition}
(Badge Threshold): A badge's \textit{threshold} denotes the minimum required \textit{contributions} required for users to get the badge. For badges in $\mathcal{B}$, their \textit{threshold} can be represented as $\mathbf{\theta} = [\theta_1, \theta_2, \cdots, \theta_m]$.
\end{definition}

For a given user $u_i$, if his/her contribution to badge $b_j$, i.e., $c_{i,j}$, is greater than $b_j$'s threshold $\theta_j$, then $u_i$ will get $b_j$, which can be represented with the following \textit{badge indicator function}:
\begin{align*}
I(c_{i,j} \ge \theta_j) = \begin{cases} 
1,  & \mbox{if }c_{i,j} \ge \theta_j, \\
0, & \mbox{otherwise.}
\end{cases}
\end{align*}
Furthermore, the badges that user $u_i$ have received can be represented as the \textit{badge indicator vector} $\mb{I}_i = [I(c_{i,1} \ge \theta_1), I(c_{i,2} \ge \theta_2), \cdots, I(c_{i,m} \ge \theta_m)]$. Before the system starts to operate and players begin to invest their efforts, the badge system designer needs to specify the badge system settings in advance, which is formally defined as the \textit{badge system mechanism} in this paper. 

\begin{definition}
(Badge System Mechanism): \textit{Badge system mechanism} describes various detailed aspects of the system. In addition to \textit{badge thresholds}, the \textit{mechanism} of badge system in online social networks can also involve the categories, number, names, IDs, levels of badges, as well as methods to get the badges, etc.
\end{definition}

\begin{figure*}[t]
\centering
\subfigure[linear value function]{\label{eg_fig_2_1}
    \begin{minipage}[l]{0.47\columnwidth}
      \centering
      \includegraphics[width=1.0\textwidth]{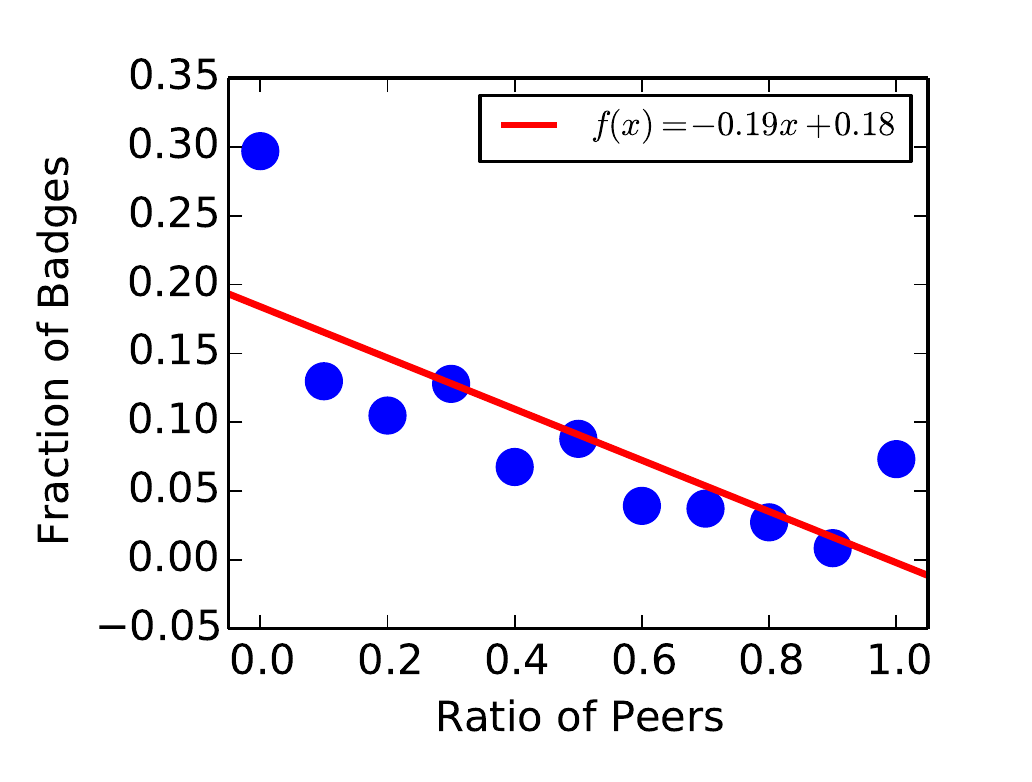}
    \end{minipage}
}
\subfigure[quadratic value function]{ \label{eg_fig_2_2}
    \begin{minipage}[l]{0.47\columnwidth}
      \centering
      \includegraphics[width=1.0\textwidth]{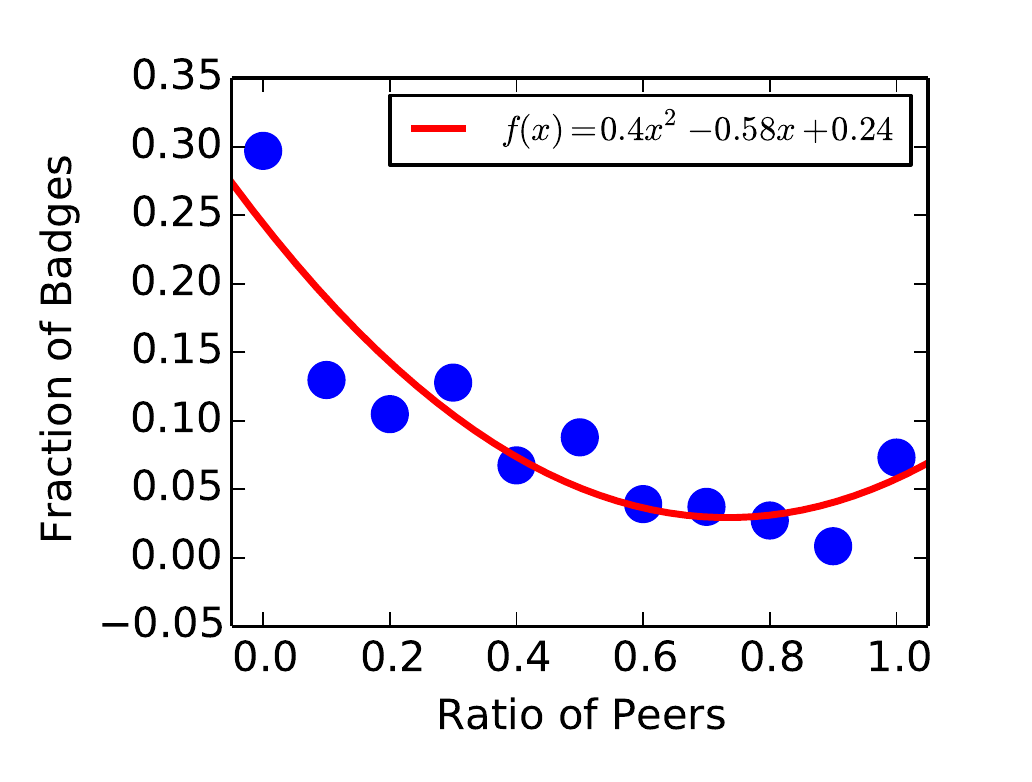}
    \end{minipage}
}
\subfigure[cubic value function]{\label{eg_fig_2_3}
    \begin{minipage}[l]{0.47\columnwidth}
      \centering
      \includegraphics[width=1.0\textwidth]{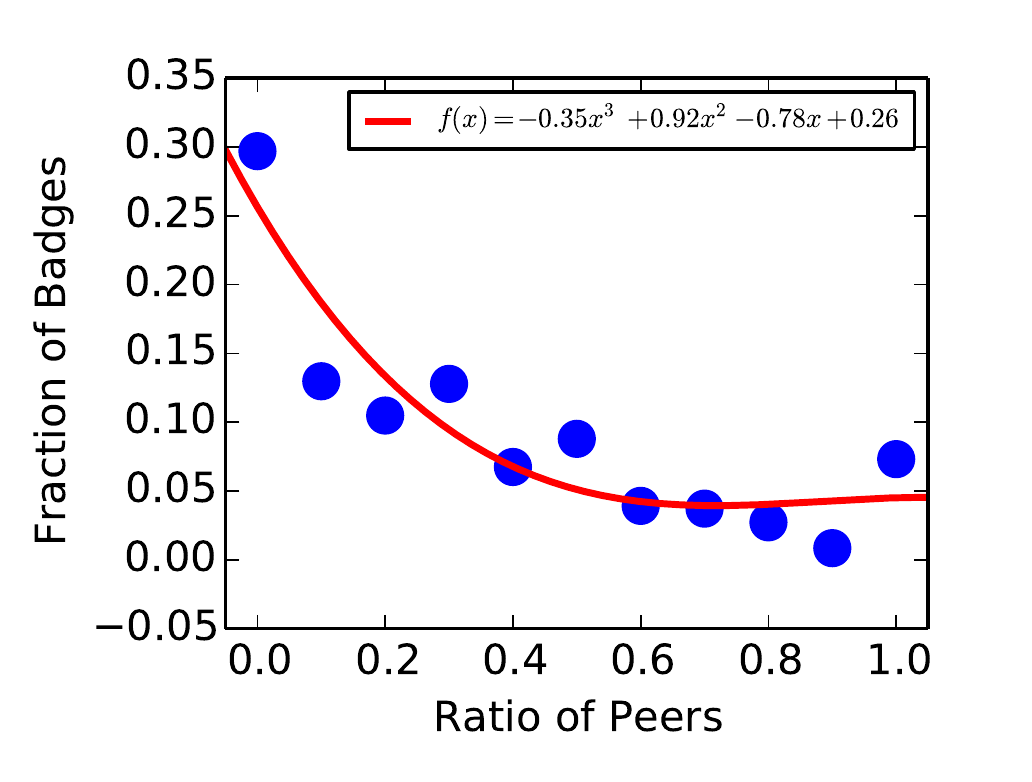}
    \end{minipage}
}
\subfigure[exp. value function]{\label{eg_fig_2_4}
    \begin{minipage}[l]{0.47\columnwidth}
      \centering
      \includegraphics[width=1.0\textwidth]{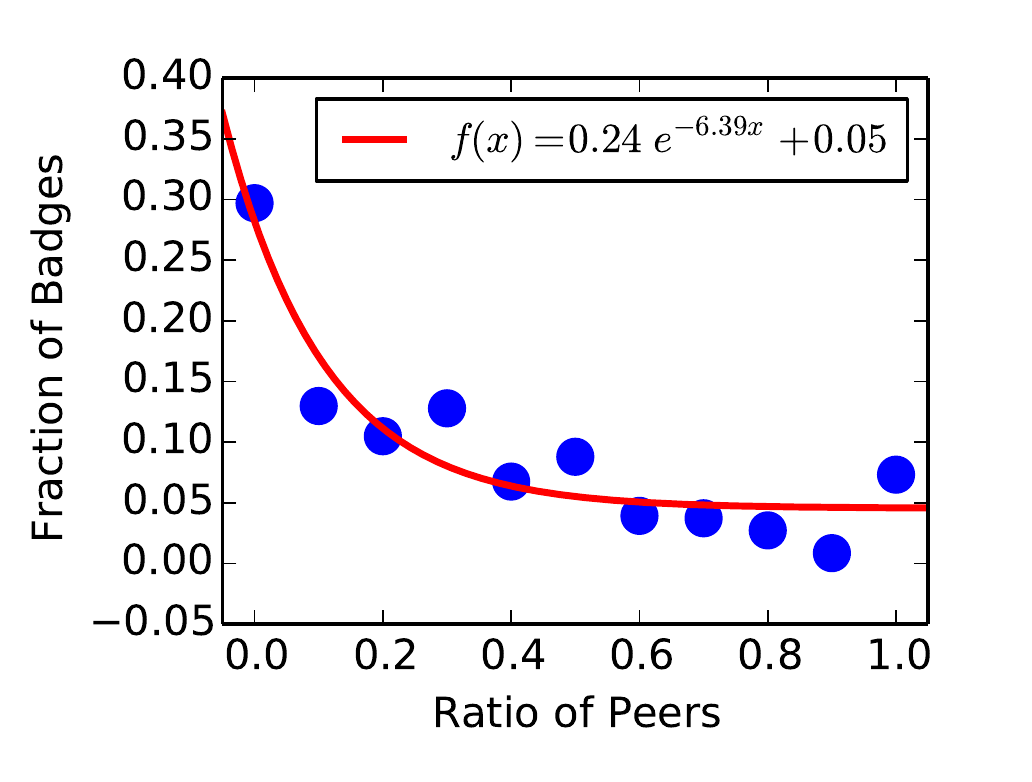}
    \end{minipage}
}
\vspace{-10pt}
\caption{Estimated value functions by fitting the data.}\label{eg_fig_2}\vspace{-12pt}
\end{figure*}

\section{Badge Value Functions} \label{sec:value}

The motivation of users being willing to devote efforts to get badges in online social networks is because these badges are valuable to them. Depending on the specific scenarios, the value of badges for users can be quite different. According to the observations in Section~\ref{sec:intro}, the effects of badges on users' social activities can be models as three different kinds of values of these badges. In this section, we will introduce three different value functions which can capture the badge values from different perspectives: (1) \textit{personal interest value function}, (2) \textit{peer leadership value function}, and (3) \textit{network trend value function}.

\subsection{Peer Leadership Value Function}\label{subsec:peer}



In our daily life, on the one hand, people want to be different from the public, but, on the other hand, they may also want to follow the mainstream as well. We have similar observations about users badge achieving activities in online social networks. Users in online social networks want to be the first to win certain badges in their communities, which can show their uniqueness and make them stand out from his/her peers. Meanwhile, if most of the peers have obtained a certain badge, users will follow their friends to get the badge to extract themselves from the backward position. To depict the effectiveness of badges to make users be either more superior to his peers or closer to other leading peers, we formally define the \textit{peer leadership badge value} in this part.

\subsubsection{Peer Leadership Value Function Definition}

\begin{definition}
(Peer Leadership Value): The \textit{peer leadership value} of badge $b_j$ for a user $u_i$ is defined as a function about the ratio of $u_i$'s peers who have obtained badge $b_j$ already. Let $\Gamma(u_i)$ be the neighbor set of user $u_i \in \mathcal{U}$, in which users who have achieved badge $b_j$ before $u_i$ can be represented as $\Psi(u_i, b_j) = \{u_m | (u_m \in \Gamma(u_i)) \land (\mb{I}_m(j) = 1) \}$. The \textit{peer leadership value} function of badge $b_j$ for user $u_i$ can be represented as function 
$$v^{ps}(u_i, b_j|\Gamma(u_i)) = f(\frac{|\Psi(u_i, b_j)|}{|\Gamma(u_i)|}), \Psi(u_i, b_j) \subset \Gamma(u_i).$$
\end{definition}

The concrete representation of the \textit{peer leadership value} functions can be quite diverse depending on the selected function $f(\cdot)$. In this paper, we try $4$ different functions, and the corresponding peer leadership value functions are listed as follows: 
\begin{itemize}
\item \textit{linear peer leadership value function}\begingroup\makeatletter\def\f@size{8}\check@mathfonts
$$v^{ps}_l(u_i, b_j|\Gamma(u_i)) = a\left(\frac{|\Psi(u_i, b_j)|}{|\Gamma(u_i)|}\right) + b,$$\endgroup
\item \textit{quadratic peer leadership value function}\begingroup\makeatletter\def\f@size{8}\check@mathfonts
\begin{align*}
&v^{ps}_p(u_i, b_j|\Gamma(u_i)) = a \left({\frac{|\Psi(u_i, b_j)|}{|\Gamma(u_i)|}}\right)^2 + b \left({\frac{|\Psi(u_i, b_j)|}{|\Gamma(u_i)|}}\right) + c,
\end{align*}\endgroup
\item \textit{cubic peer leadership value function}\begingroup\makeatletter\def\f@size{8}\check@mathfonts
\begin{align*}
v^{ps}_c(u_i, b_j|\Gamma(u_i)) &= a \left({\frac{|\Psi(u_i, b_j)|}{|\Gamma(u_i)|}}\right)^3 + b \left({\frac{|\Psi(u_i, b_j)|}{|\Gamma(u_i)|}}\right)^2 \\
&+ c \left({\frac{|\Psi(u_i, b_j)|}{|\Gamma(u_i)|}}\right) + d,
\end{align*}\endgroup
\item \textit{exponential leadership value function}\begingroup\makeatletter\def\f@size{8}\check@mathfonts
$$v^{ps}_e(u_i, b_j|\Gamma(u_i)) = a\times e^{-b \left({\frac{|\Psi(u_i, b_j)|}{|\Gamma(u_i)|}}\right)} + c,$$\endgroup
\end{itemize}
where $a$, $b$, $c$ and $d$ are the coefficients in the functions which can be learnt from the historical data.

Given the real \textit{peer leadership value function} of badges in the social network, $f(x)$, the optimal parameters can be learnt by minimizing the following objective function:
$$\hat{\mb{\omega}} = \arg \min_{\mb{\omega}}  \int_{0}^{1} |f(x) - v^{ps}(x)|\, dx$$
where $\mb{\omega}$ is the vector of coefficients (e.g., $\mb{\omega} = [a, b]$ in the linear peer leadership value function) and $x = \frac{|\Psi(u_i, b_j)|}{|\Gamma(u_i)|} \in [0,1]$. To resolve the function, we propose to learn the coefficients by fitting the $11$ discrete points shown in Figure~\ref{eg_fig_1_3} and can get different estimated \textit{peer leadership value functions} in Figures~\ref{eg_fig_2_1}-\ref{eg_fig_2_4} respectively.



\subsubsection{Peer Leadership Value Function Evaluation}\label{subsubsec:test-exp1}

\noindent \textbf{Experiment Settings}

The higher \textit{peer leadership value} a badge has, the more likely a user will try to obtain it. To test the effectiveness of the above introduced \textit{peer leadership value functions}, we conduct an experiments on the Foursquare badge system dataset introduced in Section~\ref{sec:formulation}. In the experiment, badges achieved by less than $100$ users are removed and the remaining badges achieved by users are organized in a sequence of (user, badge) pairs according to their achieving timestamps. These (user, badge) pairs are divided into two subsequences according to their relative timestamps order: the training set and testing set, the proportion of whose sizes is $9:1$. In addition, a set of non-existing (user, badge) pairs which is of the same size as the positive test set are randomly sampled from the network as the negative test set, which together with the positive test set are used to form the final testing set. Pairs in the training set are regarded as the historical data, based on which we calculate the values of pairs in the testing set and output them as the confidence scores of these pairs. 

The evaluation metrics applied in the experiment is AUC. In statistics, a receiver operating characteristic (ROC), or ROC curve, is a graphical plot that illustrates the performance of a binary classifier system as its discrimination threshold is varied. The curve is created by plotting the true positive rate (TPR) against the false positive rate (FPR) at various threshold settings. The area under the ROC curve is usually quantified as the AUC score. When using normalized units, the area under the curve (i.e.,, AUC) is equal to the probability that a classifier will rank a randomly chosen positive instance higher than a randomly chosen negative one (assuming ``positive'' ranks higher than ``negative''). Generally speaking,  larger AUC score corresponds to better performance of the prediction model.

\noindent \textbf{Experiment Results}

We learn the coefficients of different value functions with the training set and apply the learnt function to calculate the \textit{peer leadership values} of pairs in the testing set. The results are available in Figure~\ref{fig:peer_leadership_value_function}. From the results, we observe that AUC achieved by the \textit{quadratic peer leadership function} is $0.65$, which is slightly better than other value functions, and the AUC scores obtained by the \textit{linear}, \textit{cubic} and \textit{exponential peer leadership functions} are $0.58$, $0.63$, and $0.62$ respectively. Here, quadratic function can outperform cubic and exponential functions can because of the reason that cubic function may suffer from the overfitting problems a lot. Next, we will use the \textit{quadratic peer leadership function} as the only \textit{peer leadership function}, which will be compared with other value functions in Figure~\ref{fig:all_value_function}.

\subsection{Network Trend Value Function}\label{subsec:network}

Besides the effects of personal interests and peer pressure, there exists some global trend about the network steering users badge achievement activities in the whole network. In online social network, users achieve badges in a sequential time order. For example, badges achieved by user $u_i$ can be organized into a sequential transaction $\langle b^i_{1}, b^i_{2}, \cdots, b^i_{l} \rangle$ according to the achieving timestamps, where $u_i$ got badge $b^i_{p}$ before $b^i_{q}$ if $p < q$. For all users in $\mathcal{U}$, we can represent the badge achievement sequential transactions as $\{u_1: \langle b^1_{1}, b^1_{2}, ..., b^1_{l}\rangle, u_2: \langle b^2_{1}, b^2_{2}, ..., b^2_{o}\rangle, \cdots, u_n: \langle b^n_{1}, b^n_{2}, ..., b^n_{q}\rangle\}$

The network influence can be captured by extracting the frequent badge achieving sequential patterns from the transactions and many different pattern extraction methods have been proposed so far. In this paper, PrefixSpan proposed by Pei et al. \cite{PHMPCDH01} is applied. Consider, for example, we extract two frequent sequence patterns:  pattern 1: $\langle b_l, b_o, \cdots, b_p\rangle$ and pattern 2: $\langle b_l, b_o, \cdots, b_p, b_q\rangle$ with supports $support(\mbox{pattern 1})$ and $support(\mbox{pattern 2})$ respectively from the network. Rule $r$ can be generated based on pattern 1 and pattern 2 representing that for users who have obtained badges in $\langle b_l, b_o, \cdots, b_p\rangle$ has a chance of $conf$ to get badge $b_q$:
$$r: \langle b_l, b_o, \cdots, b_p\rangle \to \langle b_q\rangle, conf = \frac{support(\mbox{pattern 2})}{support(\mbox{pattern 1})},$$
where $\langle b_l, b_o, \cdots, b_p\rangle$ is called the antecedent of rule $r$ (i.e., $ant.(r)$) and $\langle b_q\rangle$ is named as the consequent  of $r$ (i.e., $con.(r)$). Score $conf(r) = \frac{support(\mbox{pattern 2})}{support(\mbox{pattern 1})}$ is called the confidence of rule $r$. Various rules together with their confidence scores can be generated based on the frequent sequence pattern mining results, which can be represented as set $\mathcal{R}$, based on which we can define the \textit{network trend value function} as follow.

\begin{definition}
(Network Trend Value Function): For a given user $u_i$, who has achieved a sequence of badges $\mathcal{H} = \langle b_{i_1}, b_{i_2}, \cdots, b_{i_{|\mathcal{H}|}}\rangle$ already, the \textit{network trend value function} of badge $b_j$ for $u_i$ is defined as the maximal confidence score of rules that can be applied to badges in $\mathcal{H}$ and badge $b_j$, i.e.,
$$v^{nt}(u_i, b_j | \mathcal{H}) = \max\{conf(r) | r \in \mathcal{R}, ant.(r) \subset \mathcal{H}, con.(r) = b_j\}.$$
\end{definition}

We evaluate the effectiveness of the introduced \textit{network trend value} of badges based on the same experiment setting introduced in Section~\ref{subsubsec:test-exp1}. As shown in Figure~\ref{fig:all_value_function}, \textit{network trend value} based badge predictor along can achieve an AUC score of $0.68$ in inferring potential badge achievement activities.

\subsection{Personal Interest Value Function}\label{subsec:interest}

Users can have their personal interests, which can steer their social activities in online social networks. For example, sport enthusiasts may visit gyms and outdoor places frequently, while gourmets tend to go to good restaurants on the other hand. Users' personal interests can be revealed from the badges obtained in the past. For example, for a given user $u_i$ who has already achieved the ``Gym Rat'' badges of levels from 1 to 4, it can show that $u_i$ can like doing sports a lot and ``Gym Rat'' of level 5 can meet his interest and can be of great value to him. Viewed in this way, the value of badges can be evaluated with the badges that users obtained in the past.


\begin{definition}
(Personal Interest Value): For a given user $u_i$ and the set of badges obtained by $u_i$ in the past, i.e., $\mathcal{H}$, the \textit{personal interest value} of badge $b_j$ for user $u_i$ is defined to be
$$v^{pi}(u_i, b_j | \mathcal{H}) = \frac{\sum_{b_k \in \mathcal{H}} s(b_j, b_k) v^{pi}(u_i, b_k)}{\left | \mathcal{H} \right |},$$
where $s(b_j, b_k)$ denotes the similarity score between badge $b_j$ and $b_k$ and $v^{pi}(u_i, b_k)$ represents the \textit{personal interest value} of badge $b_k$ for user $u_i$.
\end{definition}

For badge $b_k \in \mathcal{H}$ that $u_i$ has obtained in the past, we define the \textit{personal interest value} of $u_i$ to badge $b_j$ as $1.0$ (i.e., $v^{pi}(u_i, b_k) = 1.0$, for $\forall b_k \in \mathcal{H}$). The similarity score between any two badges, e.g., $b_j$ and $b_k$, is defined as the \textit{Jaccard's Coefficient} score \cite{LK03} of user sets who have achieved $b_j$ and $b_k$ (i.e., $\Gamma(b_j)$ and $\Gamma(b_k)$) respectively in the network:
$$s(b_j, b_k) = \frac{\left | \Gamma(b_j) \cap \Gamma(b_k) \right |}{\left | \Gamma(b_j) \cup \Gamma(b_k) \right |}.$$
Based on the above descriptions, the \textit{personal interest value} of badge $b_j$ for user $u_i$ can be represented as 
$$v^{pi}(u_i, b_j | \mathcal{H}) = \frac{\sum_{b_k \in \mathcal{H}} \frac{\left | \Gamma(b_j) \cap \Gamma(b_k) \right |}{\left | \Gamma(b_j) \cup \Gamma(b_k) \right | }}{\left | \mathcal{H} \right |}.$$

The effectiveness of the \textit{personal interest value} of badge is evaluated with a similar experiment setting, whose result is available in Figure~\ref{fig:all_value_function}. We can observe that ranking badges according to their \textit{personal interest values} for each user can achieve an AUC score of $0.66$.


\subsection{Comprehensive Value Function Definition and Evaluation}\label{subsec:combined}


To capture the information from all the three aspects in calculating badge values, we define the \textit{comprehensive value} value function as a combination of the \textit{personal interest value}, \textit{peer leadership value} and \textit{network trend value} functions: \begingroup\makeatletter\def\f@size{7}\check@mathfonts
\begin{align*}
v^{c}(u_i, b_j) = \alpha \cdot v^{pi}(u_i, b_j) + \beta \cdot v^{pp}(u_i, b_j) + (1 - \alpha - \beta) v^{ns}(u_i, b_j),
\end{align*}\endgroup
where parameters $\alpha$, $\beta$ are assigned with value $\frac{1}{3}$ for simplicity in this paper.



\begin{figure}[t]
\centering
    \begin{minipage}[l]{1.0\columnwidth}
      \centering
      \includegraphics[width=\textwidth]{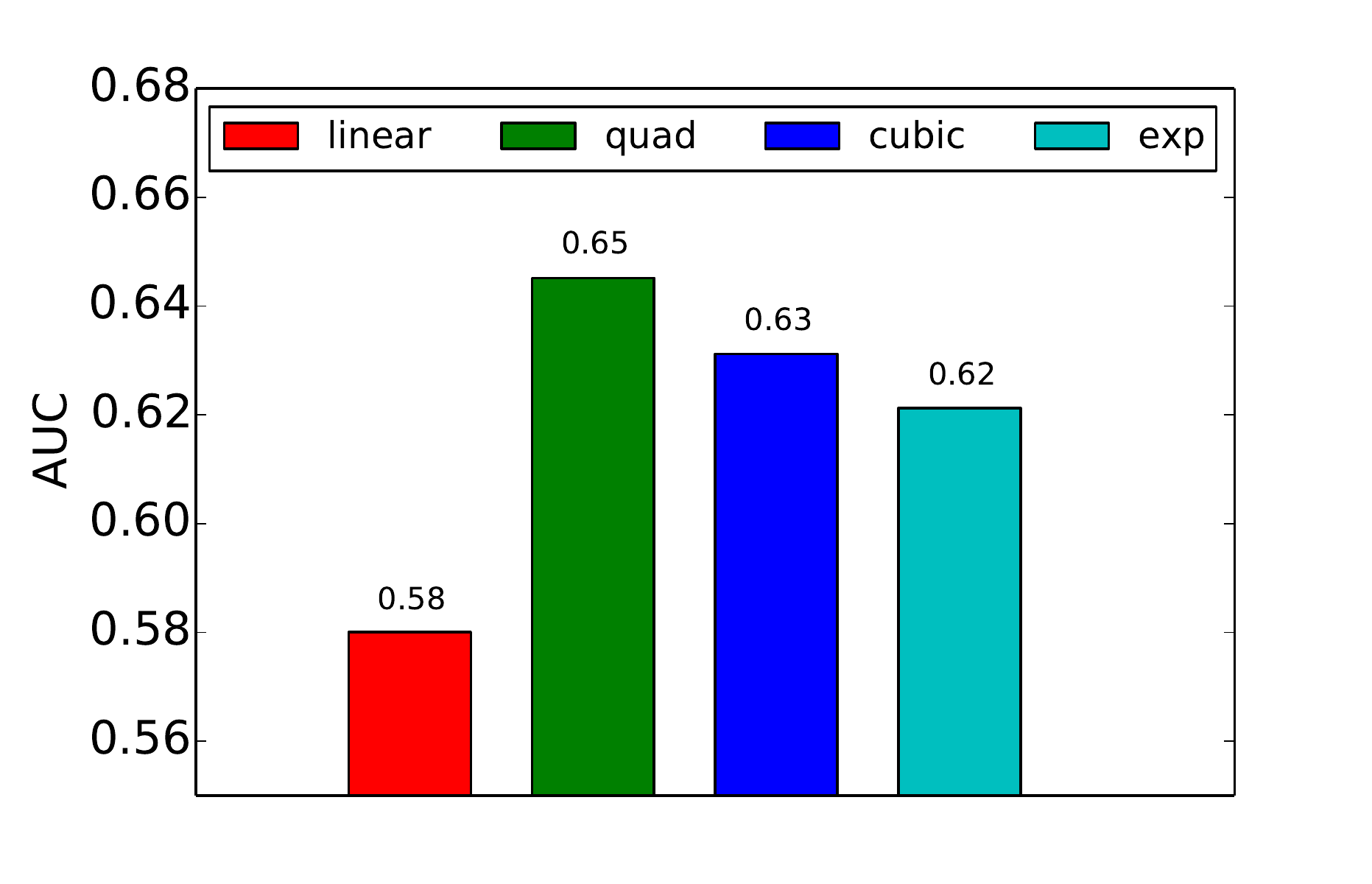}
    \end{minipage}
\vspace{-12pt}
  \caption{Comparison of peer leadership value functions}\label{fig:peer_leadership_value_function}\vspace{-12pt}
\end{figure}

\begin{figure}[t]
\centering
    \begin{minipage}[l]{1.0\columnwidth}
      \centering
      \includegraphics[width=\textwidth]{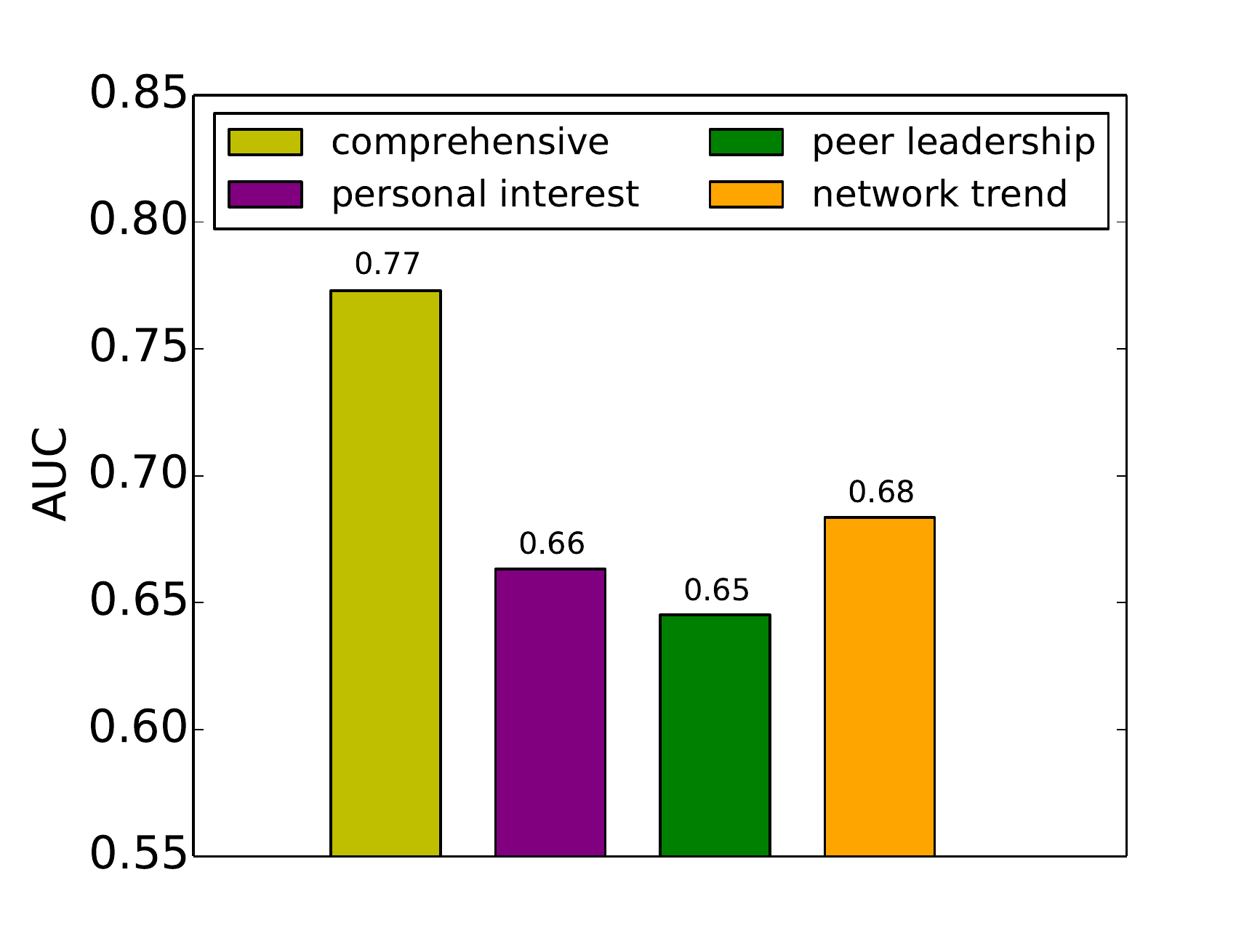}
    \end{minipage}
\vspace{-12pt}
  \caption{Comparison of comprehensive value function and other isolated value functions}\label{fig:all_value_function}\vspace{-12pt}
\end{figure}

To show the effectiveness of the above defined badge value functions in modeling users' badge obtaining activities, we also compare their performance in inferring users' badge achieving  probabilities in the Foursquare badge system dataset. For the \textit{peer leadership value function}, the \textit{quadratic} function is used as it can achieve the best performance in Figure~\ref{fig:peer_leadership_value_function}. Experiment setting here is identical to that introduced in Section~\ref{subsubsec:test-exp1} and the results is available in Figure~\ref{fig:all_value_function}. From the result we observe that \textit{network trend value function} performs better than \textit{personal interest} and \textit{peer leadership} value functions, which can achieve AUC scores about $0.68$, $0.66$, and $0.65$ respectively. Meanwhile, the \textit{comprehensive value function} that merge the isolated value functions together can improve the performance greatly and can obtain AUC score is $0.77$, which is $13.2\%$, $16.7\%$, and $18.5\%$ higher than the AUC scores achieved by \textit{personal interest}, \textit{peer leadership} and \textit{network trend} value functions.

%
%
%

\section{User Utility Function}\label{sec:utility}

Value of badges is the reward that users can receive from the system. Meanwhile, to get the reward, they also need to afford certain costs introduced when finishing the required tasks. Generally, if the reward is greater than the cost, the badge will deserve the efforts. For example, if user $u_i$ can get badge $b_j$ eventually, then $u_i$ can get a \textit{reward}, i.e., the value of $b_j$ to $u_i$ (e.g., $v^c(u_i, b_j)$) and also needs to pay some costs, i.e., his \textit{cumulative efforts} $\hat{e}_{i,j}$ devoted to $b_j$. The \textit{payoff} user $u_i$ can receive by devoting efforts to badge $b_j$ is defined to be the \textit{utility} of $b_j$ for $u_i$. The formal definitions of the \textit{reward}, \textit{cost} and \textit{utility} functions of badges for users are available as follows:

\subsection{User Utility Function Definition}

\begin{definition}
(Reward Function): The \textit{reward function} of user $u_i$ in achieving badge $b_j$ is defined as
$$reward(u_i, b_j) = I(c_{i,j} \ge \theta_j)v^c(u_i, b_j).$$
If $u_i$ can obtain $b_j$, then the \textit{reward} $u_i$ can achieve will be the \textit{comprehensive value} of badge $b_j$ for $u_i$; otherwise, the reward will be $0$.
\end{definition}

\begin{definition}
(Cost Function): To achieve a certain badge, e.g., $b_j$, the \textit{cost} that $u_i$ needs to pay is defined as the \textit{cumulative effort} that $u_i$ invests on $b_j$:
$$cost(u_i, b_j) = \hat{e}_{i,j}.$$
\end{definition}

The minimum efforts $\tilde{e}_{i,j}$ required for user $u_i$ to get badge $b_j$ is determined by $u_i$'s ability in achieving $b_j$ as well as the badge threshold of $b_j$, which can be represented as
\begin{align*}
\tilde{e}_{i,j} &= \arg \min_{\hat{e}} (a_{i,j} \hat{e}_{i,j} \ge \theta_j) = \frac{\theta_j}{a_{i,j}}
\end{align*}

\begin{definition}
(Utility Function): The \textit{utility function} of $u_i$ in achieving $b_j$ is defined as
\begin{align*}
utility(u_i, b_j) &= reward(u_i, b_j) - cost(u_i, b_j)\\
&= I(c_{i,j} \ge \theta_j)v^c(u_i, b_j) - \hat{e}_{i,j}.
\end{align*}
If $u_i$ can get $b_j$, i.e., $a_{i,j} \hat{e}_{i,j} \ge \theta_j$, then $utility(u_i, b_j) = v^c(u_i, b_j) - \hat{e}_{i,j}$; otherwise, $utility(u_i, b_j) = - \hat{e}_{i,j}$.
\end{definition}

The \textit{overall utility function} of users $u_i$ over all badges in $\mathcal{B}$ is represented as
\begin{align*}
utility(u_i) &= \sum_{b_j \in \mathcal{B}} utility(u_i, b_j)\\
&= \mb{I}_i^\top \mb{v}^c_{i} - \left\| \mb{\hat{e}}_i \right\|_1,
\end{align*}
where $\mb{I}_i^\top$ denotes the transpose of the \textit{badge indicator vector} of $u_i$, $\mb{v}^c_{i} = ({v}^c_{i,1}, {v}^c_{i,2}, \cdots, {v}^c_{i,m})$ is the \textit{comprehensive value} vector of all badges for $u_i$, $\mb{\hat{e}}_i$ is the \textit{cumulative effort vector} of $u_i$ and $\left\| \mb{\hat{e}}_i \right\|_1 = \sum_{j=1}^m \hat{e}_{i,j}$ is the $L_1$ norm of vector $\mb{\hat{e}}_i$.

\subsection{User Utility Function Evaluation}

To demonstrate the effectiveness of the utility in modeling users' badge achieving activities, we conduct the experiments to show the performance of \textit{user utility function} in inferring users badge achieving activities. Experiment settings here is identical to those introduced before but, to calculate the utilities of different badges for users, we need to know users'  total cumulative efforts, ability distributions, and badge thresholds in advance.

\noindent \textbf{Inference of Cumulative Effort}: Active users in online social networks are assumed to have more \textit{cumulative efforts}. In our dataset, the activeness measure can be defined as the number of badges users achieved. And the cumulative effort of user, e.g., $u_i$, can be obtained by normalizing the badge numbers to the range of $[0, 1]$ with equation $\frac{\#(u_i) - \#{min}}{\#{max} - \#{min}}$, where $\#(u_i)$ is the number of badges achieved by $u_i$ and $\#{max}$ and $\#{min}$ are the maximal and minimal number of badges achieved by users in $\mathcal{U}$ respectively.

\noindent \textbf{Inference of User Ability Vector}: In the training set, user $u_i$'s inferred ability vector is defined to be $\mb{a}^{infer}_i = (a_{i,1}, a_{i,2}, \cdots, a_{i,m})$ of length $m = \left | \mathcal{B} \right |$, where $a_{i,j}$ is the number of times that $u_i$ obtained badge of category $b_j$ in the training set. Each user is assumed to have the same amount of ability but can be distributed differently. Vector $\mb{a}^{infer}_i$ is normalized by the total number of achieved badges to ensure $\left | \mb{a}^{infer}_i \right |_{1} = 1$. Considering that users can have their hidden abilities, a random ability vector $\mb{a}^{random}_i$ of length $m$ is generated whose cells contain random numbers in $[0, 1]$ and $\alpha \cdot \mb{a}^{infer}_i + (1.0 - \alpha) \cdot \mb{a}^{random}_i$ is used as the final ability vector of user $u_i$. In this paper, we set parameter $\alpha = 0.85$.

\noindent \textbf{Inference of Badge Threshold}: Badges which are hard to achieve will be obtained later. For each badge $b_j \in \mathcal{B}$, we get all the users who have achieved $b_j$ from the training set: $\{u^{j}_{1}, u^{j}_2, \cdots, u^{j}_k\}$. For user $u^{j}_{i} \in \{u^{j}_{1}, u^{j}_2, \cdots, u^{j}_k\}$, we organize all the badges obtained by $u^{j}_i$ from the training set in a sequence according to their achieving timestamps, the index of $b_j$ in $u^{j}_i$'s achieved badge list is extracted to calculate $b_j$'s threshold. For example, if $u_i$ have achieved $p$ badges in all and the index of $b_j$ in the list is $q$, then the threshold of $b_j$ for $u_i$ is estimated as $\theta_{j,i} = \frac{p}{q}$. The threshold of badge $b_j$ is defined as the average of thresholds calculated for all these users: $\theta_{j} = \eta_j \frac{\sum_{o = 1}^k \theta_{j,o}}{k}$ where $\eta_j$ is a scaling parameter. Value $\eta_j$ is selected as large as possible but, at the same time, $\eta_j$ needs to ensure that for all users who have obtained badge $b_j$ in the training set (i.e., $\forall u^{j}_{i} \in \{u^{j}_{1}, u^{j}_2, \cdots, u^{j}_k\}$). When $u^{j}_{i}$ devotes all his cumulative effort to get $b_j$, $u^{j}_{i}$'s contribution can obtain $b_j$ in our model and, in other words, his contribution can exceed $\theta_{j}$.

\begin{figure}[t]
\centering
    \begin{minipage}[l]{1.0\columnwidth}
      \centering
      \includegraphics[width=\textwidth]{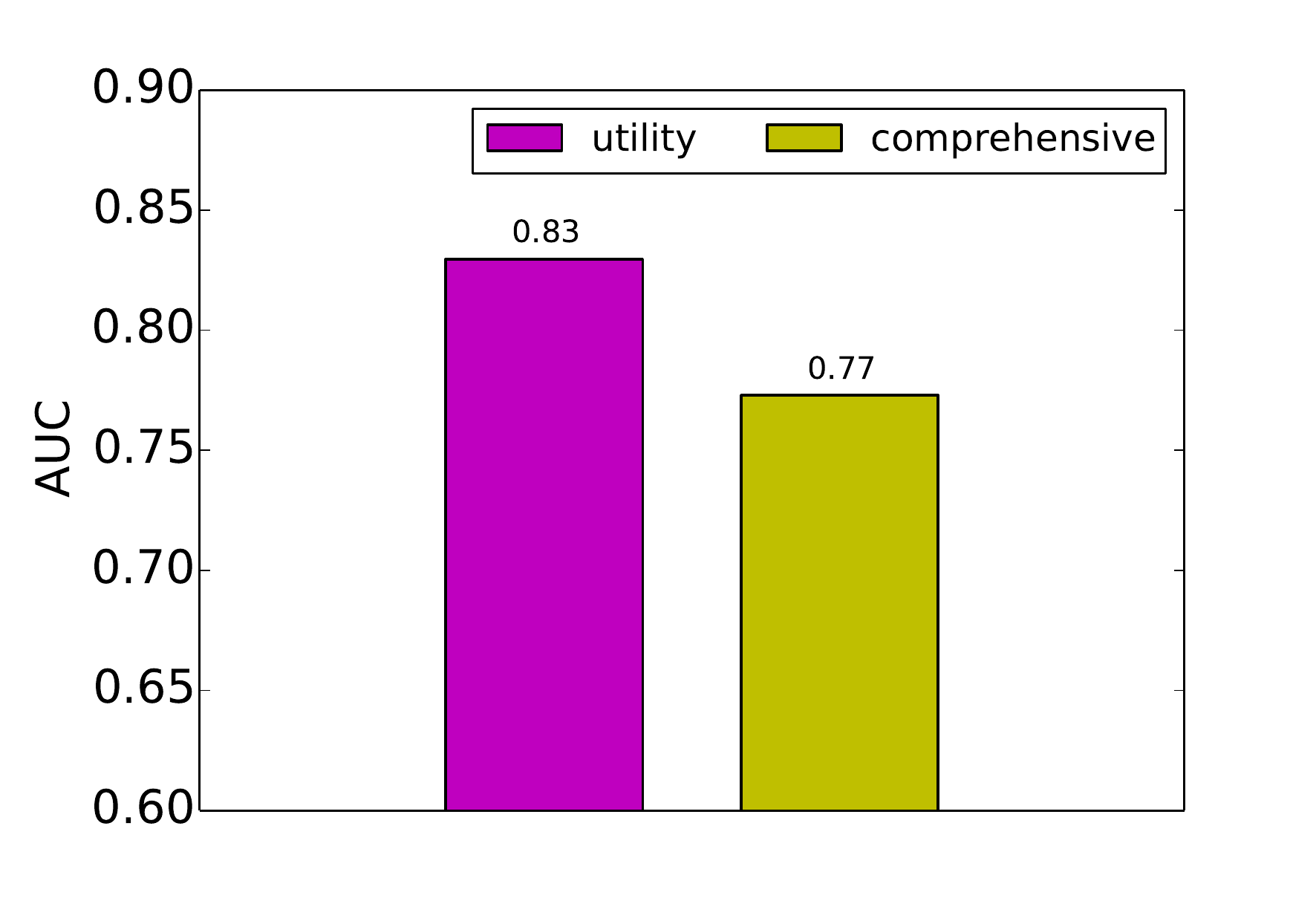}
    \end{minipage}
\vspace{-10pt}
  \caption{Comparison of utility function and comprehensive value function}\label{fig:utility_function}\vspace{-10pt}
\end{figure}

Based on the above inferred \textit{cumulative efforts}, \textit{ability} of users as well as \textit{badge thresholds}, we show the results achieved by the \textit{user utility function} in Figure~\ref{fig:utility_function} and also compare it with the \textit{comprehensive value function} introduced before. From the results, we can observe that the introduced \textit{user utility function} can perform very well in modeling users badge achieving activities. The AUC score achieved by the \textit{user utility function} is $0.83$, which is $7.8\%$ larger than the AUC score achieved by \textit{comprehensive value function} (i.e., $0.77$). As a result, \textit{user utility function} can provide a more comprehensive modeling about users' badge achievement activities.


\section{Game Among Users} \label{sec:game1}

In social networks, every user wants to maximize his/her utility in achieving badges and the value of different badges for certain user may depend on other users social activities. As a result, the badge achieving activities in online social networks can form a game among users. In traditional game theory, all the agents (e.g., users in social networks) are all assumed to be \textit{self-interested}, which means that they have their own description about the states of world they like the most and they will act in an attempt to bring about these states of the world. ``Self-interested'' doesn't necessarily mean that users tend to harm other users to maximize their payoff, as it can also include good things happening to other users as well. Meanwhile, what users can do in the \textit{game} is to determine the distribution of their \textit{cumulative efforts}, which is formally defined as the \textit{strategy} as follows.



\begin{definition}
(Strategy): A user's \textit{strategy} refers to the options that he chooses in a setting where the outcome depends not only on his own actions but also on the actions of other users. A user's \textit{strategy} can determine the actions the user will take at any stage of the game.
\end{definition}

In badge systems, users \textit{strategy} can cover various aspects of their social activities but, in this paper, we refer to the \textit{strategy} of users as the way how they distribute their \textit{cumulative efforts} for simplicity. In game theory, \textit{strategy} can be divided into two categories: (1) pure strategy, and (2) mixed strategy. The \textit{strategy} which is to select one single action in the game is referred to as the \textit{pure strategy}. In the given badge set $\mathcal{B}$, a user $u_i$ can choose to get one badge only, e.g., $b_j \in \mathcal{B}$, and devote all his/her \textit{efforts} to obtaining that badge, the \textit{strategy} of which is a \textit{pure strategy}. 

Meanwhile, let $\Pi(\mathcal{X})$ be the set of all possible distributions over set $\mathcal{X}$. Then the set of \textit{mixed strategies} for user $u_i$ is $s_i = \Pi(\mathcal{A}_i)$, where $\mathcal{A}_i$ is the set of all possible actions that $u_i$ can take. The set of all possible \textit{mixed strategy} that $u_i$ can apply is represented as set $S_i$. For simplicity, we can just regard the \textit{cumulative effort} distribution vector $\mb{\hat{e}}_i = [\hat{e}_{i,1}, \hat{e}_{i,2}, \cdots, \hat{e}_{i,m}]$ as the action distribution $\Pi(\mathcal{A}_i)$, i.e., user $u_i$'s strategy $s_i = \mb{\hat{e}}_i$.

%
%

Given the user set $\mathcal{U}$, we can represent the strategies of all users in $\mathcal{U}$ except $u_i$ as $\mb{s}_{-i} = (s_1, s_2, \cdots, s_{i-1}, s_{i+1}, \cdots, s_n)$. Thus we can write the strategies of all users in $\mathcal{U}$ as $s = (s_i, \mb{s}_{-i})$, where $s_k = \mb{\hat{e}}_k, k \in \{1, 2, \cdots, n\}$. Meanwhile, depending on users' various \textit{mixed strategies}, different kinds of social activities will be exerted in achieving badges, which can lead to different \textit{utilities}.

\begin{definition}
(Strategy Utility Function): Given user $u_i$'s and other users' strategies: $s_i$ and $\mb{s}_{-i}$, the \textit{utility} that $u_i$ can get based on $s_i$ and $\mb{s}_{-i}$ can be represented as:
\begin{align*}
u(s_i, \mb{s}_{-i}) &= utility(u_i | s_i, \mb{s}_{-i})\\
&= \sum_{j = 1}^m utility(u_i, b_j | s_i, \mb{s}_{-i}).
\end{align*}
\end{definition}

\begin{definition}
(Strategy Domination): Let $s_i$ and $s_i'$ be two \textit{mixed strategies} of user $u_i$ and $\mb{s}_{-i}$ be the strategies of all other users in $\mathcal{U}$ except $u_i$. Then, 
\begin{itemize}
\item \textit{Strict Domination}: for $u_i$, $s_i$ \textit{strictly dominates} $s_i'$ iff $u(s_i, \mb{s}_{-i}) > u(s_i', \mb{s}_{-i})$ for $\forall \mb{s}_{-i} \in S_{-i}$, where $S_{-i}$ represents the set of all potential strategies of other users except $u_i$;
\item \textit{Weak Domination}: for $u_i$, $s_i$ \textit{weakly dominates} $s_i'$ iff $u(s_i, \mb{s}_{-i}) \ge u(s_i', \mb{s}_{-i})$ $\forall \mb{s}_{-i} \in S_{-i}$ and $\exists \mb{s}_{-i} \in S_{-i}$, such that $u(s_i, \mb{s}_{-i}) > u(s_i', \mb{s}_{-i})$;
\item \textit{Very Weak Domination}: for $u_i$, $s_i$ \textit{very weakly dominates} $s_i'$ iff $u(s_i, \mb{s}_{-i}) \ge u(s_i', \mb{s}_{-i})$ for $\forall \mb{s}_{-i} \in S_{-i}.$
\end{itemize} 
\end{definition}

\begin{definition}
(Dominant Strategy): Let $s_i$ be a \textit{mixed strategy} of user $u_i$, $s_i$ is a (strictly, weakly, very weakly) \textit{dominant strategy} iff $s_i$ can (strictly, weakly, very weakly) \textit{dominate} $s_i'$ for $s_i' \in S_i, s_i' \neq s_i$, regardless of other users' strategies (i.e., $\mb{s}_{-i}$).
\end{definition}


\begin{table}[t]
\caption{Contributions of top 10 badges}
\label{tab:badge}
\centering
\begin{tabular}{ccc}
\toprule

badge name&total \# &total contributions \\
\midrule
Fresh Brew  &7878     &27.6\\
Mall Rat      &7028     &26.2\\
JetSetter     &6468     &24.5\\
Hot Tamale     &6355     &23.2\\
Great Outdoors    &5728     &21.8\\
Pizzaiolo     &4746     &17.8\\
Swimmies    &4361     &16.4\\
Bento     &3774     &13.7\\
Zoetrope     &3580     &12.9\\
Flame Broiled     &3494     &12.6\\

\bottomrule
\end{tabular}\vspace{-10pt}
\end{table}


The optimal distribution of $u_i$'s \textit{cumulative efforts} is identical to the \textit{dominant strategy} of $u_i$, which can be obtained by solving the following \textit{maximization objective function}:
$$\hat{s}_i = \arg \max_{s_i} u(s_i, \mb{s}_{-i}),$$
where $\hat{s}_i$ is the \textit{dominant strategy} of $u_i$ and other users strategies $\mb{s}_{-i} \in S_{-i}$ can take any potential value.

The above objective function is very hard to solve mathematically, as we may need to enumerate all potential strategies of all the users (including both $u_i$ and other users) in the network to obtain the global optimal strategy of $u_i$. Based on the assumption that all users are ``self-interested'', in this paper, we propose to calculate the equilibrium state of all users strategy selection process instead as follows:

We let the users to decide their optimal strategies in a random order iteratively until convergence. At first, in the $1_{st}$ round, we let users to decide their optimal strategies in a random order. For example, if we let $u_i$ be the first one to choose his  ``optimal strategy'' when other users are not involved in the system (i.e.,  $\mb{s}_{-i} = \mb{0}$), we can represent strategy selected by $u_i$'s as:
$$\tilde{s}_{i} = \arg \max_{s_i} u(s_i, \mb{0}).$$
Based on $u_i$'s ``optimal strategy'', other users in $\mathcal{U} - \{u_i\}$ (e.g., $u_j$) will take turns to decide their own ``optimal'' strategies by utilizing the selected strategies of other users. For example, let $u_j$ be the $2_{nd}$ user to decide his/her strategy right after $u_i$. The ``optimal strategy'' of $u_j$ can be represented as
$$\tilde{s}_{j} = \arg \max_{s_j} u(s_j, \{\tilde{s}_i\} \cup \mb{0}).$$
And let $u_k$ be the last user to select the ``optimal strategy'' in the $1_{st}$ round. Based on the known strategies selected by all the other users, the ``optimal strategy'' of $u_k$ can be represented as
$$\tilde{s}_{k} = \arg \max_{s_k} u(s_k, \{\tilde{s}_1, \tilde{s}_2, \cdots, \tilde{s}_{k-1}, \tilde{s}_{k+1}, \cdots, \tilde{s}_{|\mathcal{U}|}\}).$$

After finishing the $1_{st}$ round, we will start the $2_{nd}$ round and all users will decide their strategies in a random order. Such a process will continue until all users' ``optimal strategies'' selected in round $k$ is identical to those in round $k-1$ (i.e., the stationary state), which will be outputted as the final optimal strategies of all users.

%


\section{Badge System Design} \label{sec:game2}


In addition to the game among users, there also exists a game between users and badge system designer. Users in online social networks want to maximize their utilities with as few efforts as possible. Meanwhile, badge system designer who decides the badge system mechanism aims at maximizing all users contributions to the network on the other hand. In this section, we will study how to determine the optimal badge system mechanism, and provide detailed simulation analysis about the designed badge system based on the model of user badge achievement activities learnt from the previous sections.


For simplicity, we define the amount of contribution attracted by a badge as the contribution of the badge as follows:
\begin{definition}
(Badge Contribution): For a given badge mechanism $\mathcal{M}$, where the placed badge set is $\mathcal{B}$, the contribution of badge $b_j \in \mathcal{B}$ is defined as the total amount of contributions that users devoted to getting $b_j$. Based on the optimal strategy $\hat{{s}}_{i}$ of user $u_i$ obtained from the game objective function proposed in the previous section, badge $b_j$'s contribution can be represented as
$$c(b_j | \mathcal{M}) = \sum_{u_i \in \mathcal{U}} a_{i,j} \hat{s}_{i,j},$$
where $\hat{s}_{i,j}$ is the optimal strategy $u_i$ selected to get badge $b_j$.

\end{definition}

Furthermore, for a given badge mechanism $\mathcal{M}$, where the badge set is $\mathcal{B}$, the contribution of badge set $\mathcal{B}' \subset \mathcal{B}$ selected in badge mechanism $\mathcal{M}$ can be represented as
\begin{align*}
c(\mathcal{B}' | \mathcal{M}) &= \sum_{b_j \in \mathcal{B}'}\sum_{u_i \in \mathcal{U}} a_{i,j} \hat{s}_{i,j}.
\end{align*}


Different badge system mechanisms can lead to different amounts of contributions from users and the optimal one that can attract the maximum contribution is defined as the \textit{dominant badge mechanism}.

\begin{definition}
(Dominant Badge Mechanism): Given a \textit{badge mechanism} $\mathcal{M}$, in which the placed badge set is $\mathcal{B}$, if $\mathcal{M}$ can lead to more user contributions to the system than all the other \textit{badge mechanisms}, then $\mathcal{M}$ is defined as the \textit{dominant badge mechanism}:
$$\hat{\mathcal{M}} = \arg \max_{\mathcal{M}} c(\mathcal{M}) = \arg \max_{\mathcal{M}} c(\mathcal{B} | \mathcal{M}).$$
\end{definition}

\textit{Badge Mechanism} can cover lots of different aspects, e.g., badge categories, the total number of badges, tasks required to get these badges, etc. In the following parts, we will analyze these aspects one by one.


\begin{figure}[t]
\centering
    \begin{minipage}[l]{1.0\columnwidth}
      \centering
      \includegraphics[width=\textwidth]{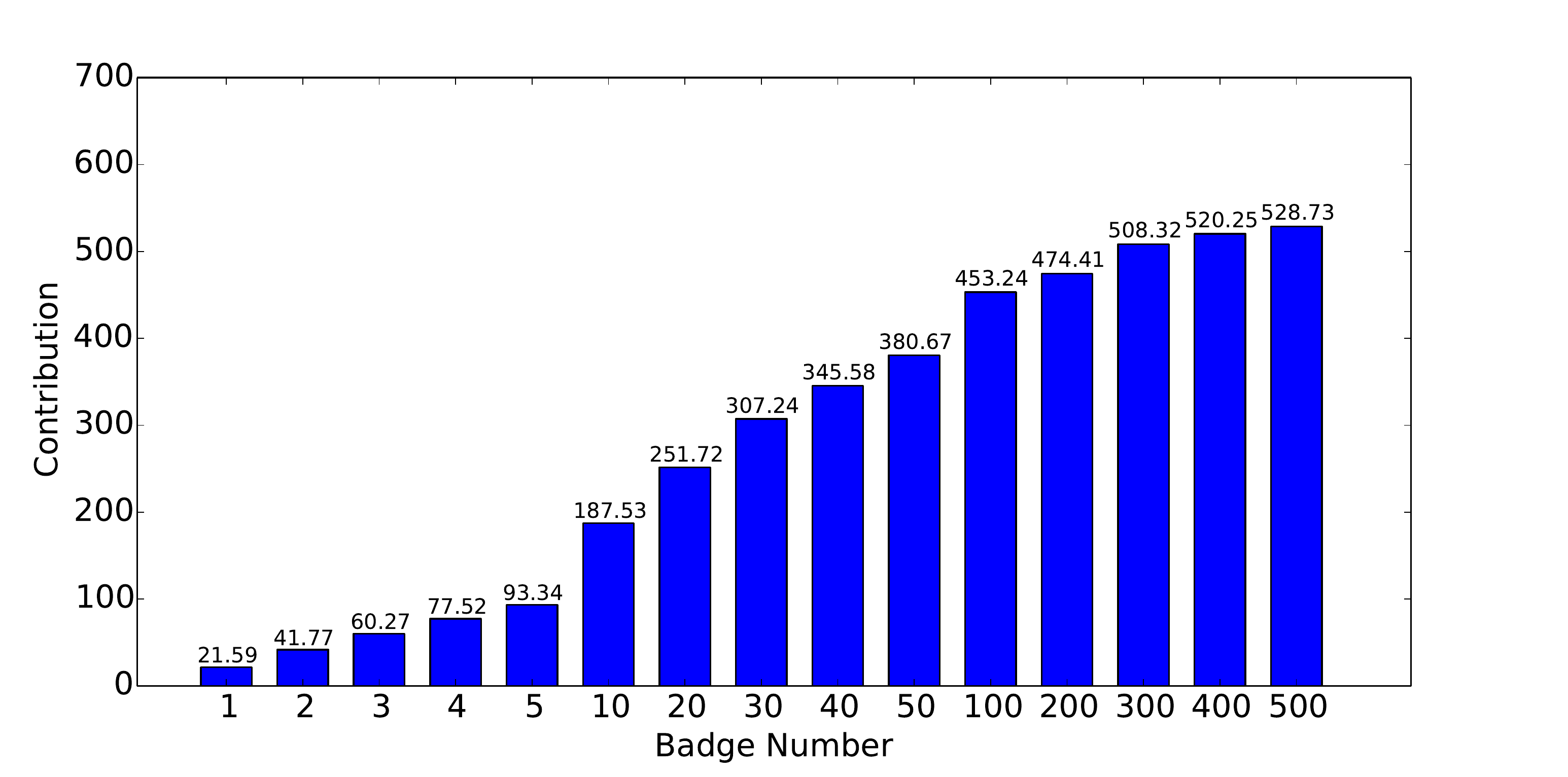}
    \end{minipage}
\vspace{-10pt}
  \caption{Simulation analysis of contributions achieved by badge mechanisms containing different numbers of badges}\label{fig:contribution_number}\vspace{-10pt}
\end{figure}

\begin{definition}
(Dominant Badge Category) For the given badge mechanism $\mathcal{M}$ (in which the badge set is $\mathcal{B}$), the \textit{dominant badge category} is defined as the badge category that can lead to the maximal contribution to the social network,
\begin{align*}
\hat{b}_j &= \arg \max_{b_j \in \mathcal{B}} c(b_j | \mathcal{M}).
\end{align*}
\end{definition}

According to the experiment settings introduced in Section~\ref{sec:utility}, users decide their optimal strategies with methods proposed in the previous section. Based on users' optimal strategies, the top $10$ most \textit{badge categories} that can lead to the maximum contribution are available in Table~\ref{tab:badge}. In the table, we show the badge names, number of times that users have obtain certain kind of badges and the contributions of each badge. Generally, popular badges (i.e., badges achieved by many users) can attract more contributions from users according to the results.

\begin{definition}
(Dominant Badge Category Set) For the given badge mechanism $\mathcal{M}$ and badge set $\mathcal{B}$, the \textit{dominant badge category set} of size $K$ is defined as the badge subset $\mathcal{B}$', $\left | \mathcal{B}' \right| = K$, which can lead to the maximal contribution to the social network,
\begin{align*}
\hat{\mathcal{B}}' &= \arg \max_{\mathcal{B}' \subset \mathcal{B}, \left | \mathcal{B}' \right| = K} c(\mathcal{B}' | \mathcal{M}).
\end{align*}
\end{definition}

To analyze the effects of badge numbers on the global contributions of badge system, we select the top $K$, $K \in \{1, 2,\cdots, 5, 10, 20,\\ \cdots, 50, 100, 200,\cdots, 500\}$ categories of badges from the network and calculate the contribution of these badges. The simulation results are given in Figure~\ref{fig:contribution_number}, where the total contribution of all these top $K$ badges will increase as $K$ increases, but the speed of the growth will slow down when $K$ is large enough (e.g., $K \ge 100$). It supports that small number of badges are already enough to approximate the optimal revenue as introduced in \cite{ISS13}.

%

Another key factor in badge system design is the \textit{badge thresholds} and the optimal badge thresholds is formally defined as the \textit{dominant badge threshold} as follows.

\begin{definition}
(Dominant Badge Threshold) For the given badge mechanism $\mathcal{M}$ and badge set $\mathcal{B}$, the \textit{dominant badge threshold} $\mb{\hat{\theta}}$ is defined as
\begin{align*}
\mb{\hat{\theta}} &= \arg \max_{\mb{{\theta}}} c(\mathcal{B} | \mathcal{M}, \mb{\theta}).
\end{align*}
\end{definition}

\begin{figure}[t]
\centering
    \begin{minipage}[l]{1.0\columnwidth}
      \centering
      \includegraphics[width=\textwidth]{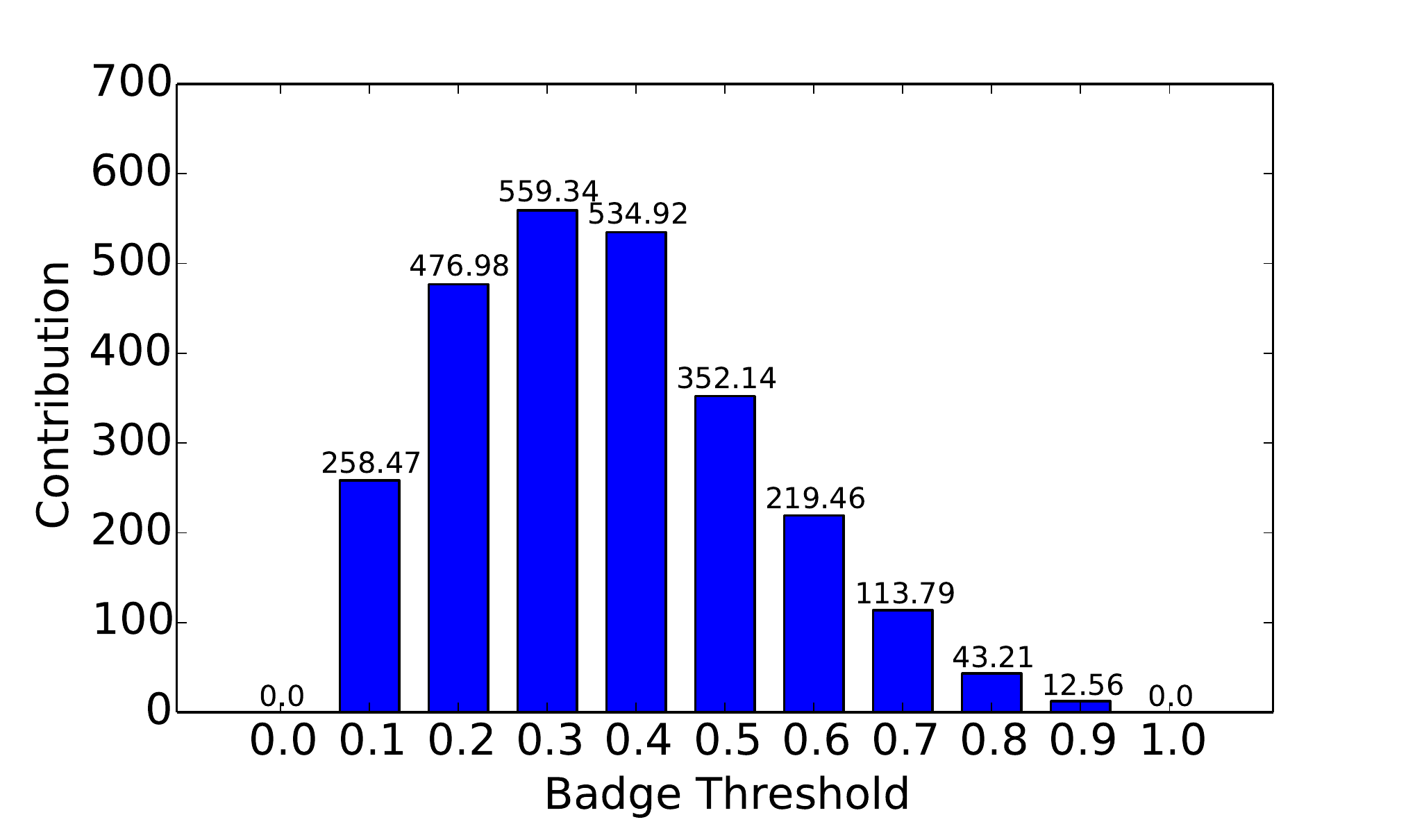}
    \end{minipage}
\vspace{-8pt}
  \caption{Simulation analysis of contributions achieved by badge mechanisms of different badge thresholds}\label{fig:contribution_threshold}\vspace{-8pt}
\end{figure}

We also study the effects of badge thresholds on the overall contributions of these badges to the network and the simulation analysis results are given in Figure~\ref{fig:contribution_threshold}. To simplify the experiment setting, we set the thresholds of all badges with the same value in $\{0.0, 0.1, \cdots, 0.9, 1.0\}$ and get the contributions obtained by the badges. As shown in Figure~\ref{fig:contribution_threshold}, the contribution of all these badges is $0$ when the threshold is $0$ and $1.0$, where threshold $0$ denotes that users can get badges without paying any efforts; threshold $1.0$ means that users need to devote all their efforts on getting the badge corresponding to the area that all their ability lies in (i.e., ability in this area is $1.0$). Contributions made to the network will increases fast as badge threshold increase at the beginning and can achieve the maximal contribution when threshold is $0.3$ and $0.4$ and then it will decreases.

\section{Related Work} \label{sec:relatedwork}

Reward systems, e.g., badge system, have been widely employed in online social networks, like Foursquare \cite{AC11, Foursquare, CSPE13}. Antin et. al. study the badges in online social networks from a social psychological perspective and give some basic introduction of badges in Foursquare \cite{AC11}. Large amount of badges are placed in Foursquare and a complete list of Foursquare badges is available \cite{Foursquare}. To obtain badges in Foursquare, users need to reveal their locations by checking in at certain locations. Carbunar et. al. study the problem between privacy preservation and badge achievement in Foursquare \cite{CSPE13}.

Users are all assumed to be ``selfish'' and want to maximize their payoff, which will form a game among users in online social networks to compete with each other. There has been a growing literature on analyzing the game among users in online social networks. Ghosh et. al. \cite{GM11, GH11, GH13} provide a game-theoretic model within which to study the problem of incentivizing high quality user generated content, in which contributors are strategic and motivated by exposure. Jain et. al. \cite{JP13} present a simple game-theoretic model of the ESP game and characterize the equilibrium behavior in their model. Their equilibrium analysis supports the fact that users appear to be coordinating on low effort words.

To achieve the maximal contribution to the sites, many works have been done on designing the badge system for online social networks. Jain et. al. \cite{JCP09} study the problem of incentive design for online question and answer sites. Anderson et. al. \cite{AHKL13} study how badges can influence and steer users behavior on social networks, which can lead both to increased participation and to changes in the mix of activities a user pursues in the network. Ghosh et. al. \cite{GH12} study the problem of implementing a mechanism which can lead to optimal outcomes in social computing based on a game-theoretic approach. Immorlica et. al. \cite{ISS13} study the badge system design whose goal is to maximize contributions. Easley et. al. \cite{FG13} take a game-theoretic approach to badge design, analyzing the incentives created by badges and potential contributors as well as their contribution to the sites.

The \textit{badge system analysis and design} problem studied in this paper is a novel problem and different from existing works on reward system analysis: (1) ``steering user behavior with badges'' \cite{AHKL13}, which studies the incentives of badges in guiding users online activities without considering the effects of social connections among users; (2) ``social status and the design of optimal badges'' \cite{ISS13}, which provides theoretical derivations of the optimal badge system design problem but fails to consider the game among users and the game between users and badge system designer; and (3) ``implementing optimal outcomes in social computing: a game-theoretic approach'' \cite{GH12}, which tries to use a game theory based method to analyze the motivations of users in getting badges but doesn't consider the ``badge system design'' problem. 
\section{Conclusion}\label{sec:conclusion}
In this paper, we study the badge system analysis and design problem, which covers (1) badge system analysis and (2) badge system design problem. We introduce the three different categories of badges value functions for users in online social networks. To depict users' payoff by achieving badges in online social networks, we formally define the utility function for users. We solve the ``badge system analysis'' problem as a game among users in social network and address the ``badge system design'' problem as a game between badge system designer and the users. Experiments conducted on real-world badge system dataset demonstrate that our model can capture users' motivations in achieving badges online very well and design badge system mechanism that can lead to maximal contributions.

\balance
\bibliographystyle{plain}
\bibliography{reference}

\end{document}